\documentclass[prl,10pt,twocolumn,superscriptaddress,showpacs]{revtex4-1}

\usepackage{epsfig}
\usepackage{amsmath}
\usepackage{graphicx}
\usepackage{graphics}
\usepackage{float}
\usepackage{amssymb}
\usepackage[usenames,dvipsnames]{color}
\usepackage{natbib}
\usepackage{epstopdf}
\usepackage{verbatim}
\usepackage{wrapfig}
\usepackage{subfigure}

\usepackage{mathtools}

\DeclarePairedDelimiter\floor{\lfloor}{\rfloor}

\newcommand{\bra}[1]{\left\langle #1\right|}
\newcommand{\ket}[1]{\left| #1\right\rangle}

\newcommand{\jn}[1]{~{\bf #1}}
\newcommand{\opav}[3]{\langle #1 | #2 | #3 \rangle}

\newcommand{\beq}{\begin{equation}}
\newcommand{\eeq}{\end{equation}}

\begin{document}

\title{Detecting weak magnetic fields using nitrogen-vacancy centers}

\author{Adam Zaman Chaudhry}
\email{adamzaman@gmail.com}
\affiliation{Department of Electrical and Computer Engineering, National University of Singapore, 4 Engineering Drive 3, Singapore 117583}
\affiliation{School of Science \& Engineering, Lahore University of Management Sciences (LUMS), Opposite Sector U, D.H.A, Lahore 54792, Pakistan}

\begin{abstract}

We show how nitrogen-vacancy centers can be used to `detect' magnetic fields, that is, to find out whether a magnetic field, about which we may not have complete information, is actually present or not. The solution to this problem comes from quantum state discrimination theory. The effect of decoherence is taken into account to optimize the time over which the nitrogen-vacancy center is allowed to interact with the magnetic field before making a measurement. We also find the optimum measurement that should be performed. We then show how multiple measurements reduce the error in detecting the magnetic field. Finally, a major limitation of the measurement process, namely limited photon detection efficiency, is taken into account. Our proposals should be implementable with current experimental technology.

\end{abstract}

\date{\today}

\maketitle

\textit{Introduction.} The ability to measure weak magnetic fields is an important problem with applications in many different fields like data storage, biomedical sciences, material science and quantum control \cite{FreemanScience2001, Budkermagnetometry, GreenbergRevModPhys1998, GrinoldsNatPhys2011}. In this regard, nitrogen vacancy (NV) defect centers \cite{JelezkoApplPhys2002, DohertyPhysRep2013}, due to their small size, impressive magnetic field sensitivity, robustness in a variety of environments, and wide-temperature range operation, have attracted widespread attention in order to reconstruct the temporal profile of an unknown weak magnetic field \cite{TaylorNatPhys2008, MazeNature2008, BalasubramanianNature2008, ChangNatNano2008, BalasubramanianNatNano2009, HallPRL2009, LaraouiAPL2010, McguinnessNatNano2011, deLangePRL2011, KagamiPhysica2011, HorowitzPNAS2012, HirosePRA2012, PhamPRB2012, NusranPRB2013, LoretzPRL2013, LeSageNature2013, GeiselmannNatureNano2013, MagesanPRA2013, CooperNatCommun2014, NusranPRB2014, JensenPRL2014,ChaudhryPRA2014NV}. The basic idea behind the use of NV centers to measure magnetic fields is to initially prepare a superposition state of two different energy levels \cite{HallMRS2013, HongMRS2013, RondinRepProgPhys2014}. In the presence of a magnetic field, a phase difference develops between the energy levels. This phase difference, which depends on the magnetic field, can then be read out in order to infer the magnetic field. 

In this paper, our emphasis is somewhat different. We do not wish to measure the parameters of an unknown magnetic field. Rather, we want to find out whether or not a magnetic field, about which we have some prior information, is actually present or not. We refer to this problem as the `detection' of the magnetic field. One can easily envisage that the ability to answer such a question can be of great use in areas such as data storage and magnetic resonance imaging. Our basic idea is to again prepare the NV center in a superposition state of two energy levels $\ket{0}$ and $\ket{1}$, defined as $\sigma_z \ket{0} = \ket{0}$ and $\sigma_z \ket{1} = -\ket{1}$ for the Pauli matrix $\sigma_z$. That is, we prepare the state $\rho_0 = \frac{1}{2}\left( \ket{0}\bra{0} + \ket{1}\bra{1} + \ket{0}\bra{1} + \ket{1}\bra{0} \right)$. The Hamiltonian for interaction of the NV center with the magnetic field $B(t)$ is $H(t) = \pi \gamma B(t) \sigma_z$ with $\gamma = 28 \, \text{Hz}/\text{nT}$. Therefore, ignoring decoherence, if no magnetic field is present, then after time $T$ the state of the NV center is still $\rho_0$, while if there is a magnetic field, the state becomes $\rho_1 = \frac{1}{2}\left(\ket{0}\bra{0} + \ket{1}\bra{1} + e^{-i\theta}\ket{0}\bra{1} + e^{i\theta} \ket{1}\bra{0} \right)$ with $\theta$ dependent on the magnetic field. Consequently, the question of whether or not there is a magnetic field becomes a problem of discriminating between the two quantum states $\rho_0$ and $\rho_1$ \cite{Helstrombook, Holevobook, HerzogJOpt2004, BergouOptics2010}, towards which we now turn. 

\textit{Quantum state discrimination.} Suppose that we are given two states $\rho_0$ and $\rho_1$, with prior probabilities $P_0$ and $P_1$ respectively, that need to be distinguished by performing a measurement that is described, in general, by a positive-operator valued measure (POVM). The POVM elements $\Pi_0$ and $\Pi_1$ are such that if $\Pi_0$ `clicks', we say that the state is $\rho_0$ and vice versa. In general, since the two states need not be orthogonal, we have a non-zero error probability $P_e = P_0 \text{Tr}(\rho_0 \Pi_1) + P_1 \text{Tr}(\rho_1 \Pi_0)$ that we choose the wrong state if we insist on making a choice between the two states. The idea then is to construct a POVM so as to minimize the error probability $P_e$. Fortunately for us, this problem has a general solution when we need to discriminate between two states \cite{Helstrombook, BergouOptics2010}. We first define the Hermitian operator $\Lambda = P_1 \rho_1 - P_0 \rho_0$ living in a Hilbert space of dimension $D_S$. We can then find the eigenvalues $\lambda_k$ of $\Lambda$, and categorize them as $\lambda_k < 0$ for $1 \leq k < k_0$, $\lambda_k > 0$ for $k_0 \leq k \leq D$ and $\lambda_k = 0$ for $D < k \leq D_S$, thereby defining $k_0$ and $D$. It can then be shown that the minimum error probability is given by $P_e = \frac{1}{2}(1 - \sum_k |\lambda_k|)$. The optimal POVM is a projective measurement given by $\Pi_0 = \sum_{k = 1}^{k_0 - 1} \ket{\phi_k}\bra{\phi_k}$ and $\Pi_1 = \sum_{k = k_0}^{D_S} \ket{\phi_k}\bra{\phi_k}$ where $\ket{\phi_k}$ are the eigenstates of $\Lambda$.

\begin{figure}[t]
\centering
\mbox{\hspace{-3mm}\subfigure{\includegraphics[scale = 0.4]{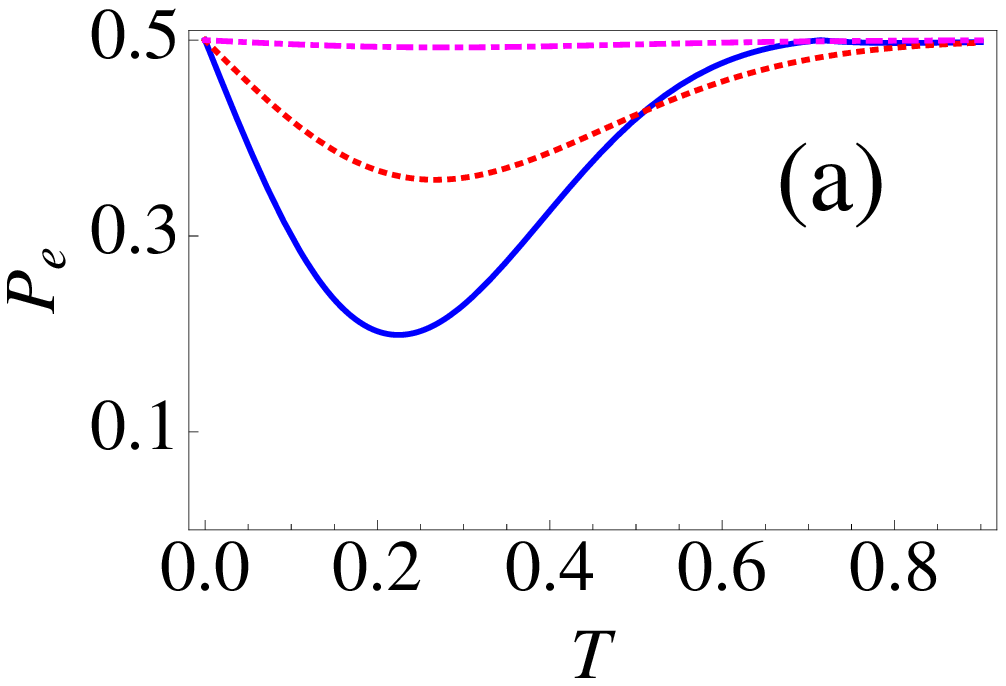}\label{DCfielddetection} }\hspace{-2mm}
\subfigure{\includegraphics[scale = 0.4]{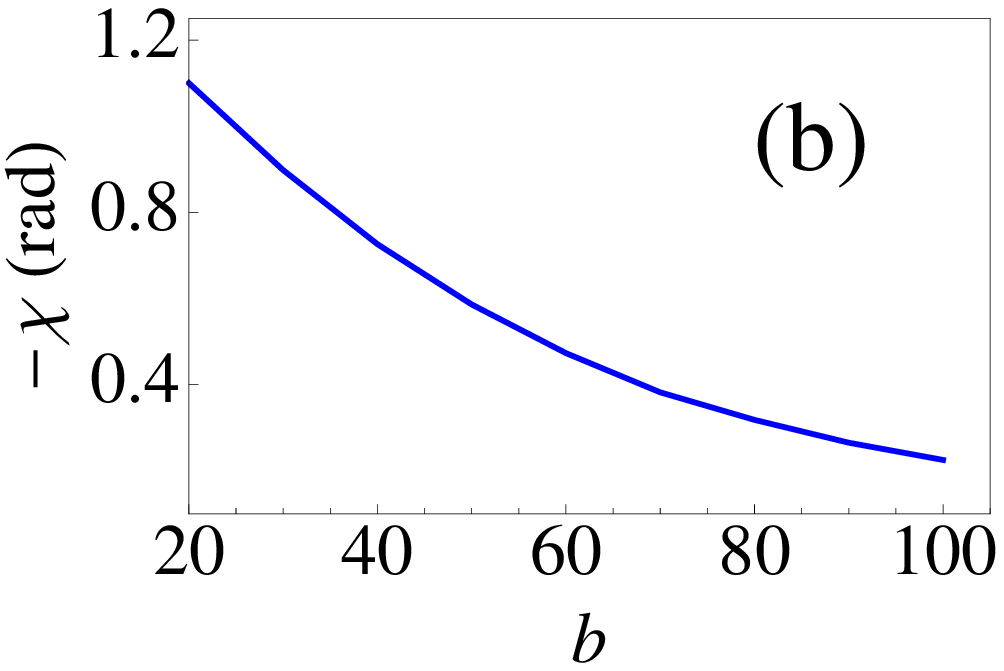}\label{DCmeasurement} }}
\caption{(color online) (a) Error probability $P_e$ as a function of the total time $T$ (in $\mu$s) for magnetic fields of strengths $1 \, \mu\text{T}$ (magenta, dot-dashed line), $20 \, \mu\text{T}$ (red, dotted line) and $50 \, \mu\text{T}$ (blue, solid line). Throughout this paper, we use $\kappa = 3.6 \, \mu s^{-1}$ and $\tau_c = 25 \, \mu s$ \cite{deLangePRL2011, WanganddeLangePRB2012}. This corresponds to a dephasing time of around $2.8$ $\mu$s. Much longer dephasing times have been obtained experimentally in ultrapure diamond samples \cite{RondinRepProgPhys2014}, allowing the detection of weaker magnetic fields. For all the numerical examples, we also use $P_0 = P_1 = 1/2$. $P_e \approx 0.2$ then means that we more than halve the error in detecting the magnetic field. (b) Variation of the optimal measurement to be performed, as quantified by $\chi$, with changing magnetic field strength $b$ (in $\mu\text{T}$).}
\label{firstfigure}
\end{figure}

Let us now apply this formalism to our case, but before doing so, we consider the effect of the environment of the NV center on the evolution of the NV center state. The NV center interacts with a spin bath composed mainly of the surrounding nitrogen defects (P1 centers). These dipolarly coupled P1 centers lead to dephasing of the NV center and negligible relaxation due to the large energy mismatch between the NV center and the P1 centers. Calculating the exact dynamics of the NV center in the presence of P1 centers is a very challenging problem because the P1 centers also interact amongst themselves. In order to make the problem tractable, a common approach is to approximate the effect of the P1 centers on the NV centers via a classical Gaussian noise field $B_d(t)$ with zero mean and correlation function $\langle B_d(0) B_d(t) \rangle = \kappa^2 e^{-|t|/\tau_c}$, where $\tau_c$ is the correlation time and $\kappa$ describes the interaction strength between the NV center and the spin bath of the P1 centers. It has been shown that such a description of the effect of the environment of the NV center describes experimental results very well \cite{WanganddeLangePRB2012}. 
Since the NV center undergoes only dephasing, it is clear that if there is no magnetic field, the NV center state becomes, after a time $T$, $\rho_0 = \frac{1}{2}\left(\ket{0}\bra{0} + \ket{1}\bra{1} + \nu \ket{0}\bra{1} + \nu \ket{1}\bra{0} \right)$ where the factor $\nu$, which depends on $T$, takes into account the effect of decoherence. On the other hand, if there is a magnetic field present, the NV center state becomes $\rho_1 = \frac{1}{2}\left(\ket{0}\bra{0} + \ket{1}\bra{1} + \nu \mu \ket{0}\bra{1} + \nu \mu^* \ket{1}\bra{0} \right)$, where the factor $\mu$ takes into account the evolution of the NV center state induced by the presence of the magnetic field that we are trying to detect. Particular forms of $\nu$ and $\mu$ will be worked out later. We thus need to discriminate between the states $\rho_0$ and $\rho_1$. Assuming that the prior probabilities for the absence and presence of the magnetic field are $P_0$ and $P_1$ respectively, we find that the eigenvalues of the operator $\Lambda = P_1 \rho_1 - P_0 \rho_0$ are 
\begin{equation*}
\lambda_{\pm} = \frac{1}{2} (P_1 - P_0) \pm \frac{\nu}{2} \sqrt{P_0^2 + P_1^2 |\mu|^2 - 2P_0 P_1 \text{Re}(\mu)}.
\end{equation*}
We can then have three cases \footnote{Strictly speaking, there are three other cases: one eigenvalue positive and the other zero, one eigenvalue negative and the other zero, and both eigenvalues zero. It is easy to see that in these cases, we can make a decision without performing any measurement.}. First, $\lambda_+$ and $\lambda_-$ are both positive. In this case, we need not perform any measurement - we can always say that there is a magnetic field. This can happen, for instance, if $P_1 > P_0$ and decoherence is very significant. Second, if $\lambda_+$ and $\lambda_-$ are both negative, we decide that there is no magnetic field. Finally, for the case that one eigenvalue is positive, while the other is negative, we find that  
\begin{equation}
P_e = \frac{1}{2}\left(1 - \nu \sqrt{P_0^2 + P_1^2 |\mu|^2 - 2P_0 P_1 \text{Re}(\mu)}\right).
\end{equation}
This is obviously the most interesting case, and the one that we will be concentrating on. In particular, if we have no previous knowledge of whether or not there is a magnetic field, then $P_0 = P_1 = 1/2$. In this case,
\begin{equation}
P_e = \frac{1}{2}\left[1 - \frac{\nu}{2}\sqrt{1 + |\mu|^2 - 2 \text{Re}(\mu)}\right].
\end{equation}
To find the optimal measurement, we find the eigenvectors of $\Lambda$ to obtain $\Pi_1 = \ket{\phi_+}\bra{\phi_+}$ and $\Pi_0 = \ket{\phi_-}\bra{\phi_-}$ with 

\[ \ket{\phi_{\pm}} = \frac{1}{\sqrt{2}}\left( \begin{array}{c}
1 \\
\mp e^{i\chi} \end{array} \right), \]
where
\begin{equation}
\tan \chi = \frac{\text{Im}(\mu)}{P_0 - P_1 \text{Re}(\mu)}.
\end{equation}

\textit{DC field detection.} As an illustration of our formalism, we start with the problem of detecting DC magnetic fields, that is $B(t) = b$. In the absence of any control fields to suppress decoherence, we find that $\nu = \langle \exp (-i \int_0^T B_d(t) \, dt )\rangle$ where $\langle \hdots \rangle$ indicates an average over the different noise realizations. Using the properties of the noise field $B_d(t)$, we find $\nu = e^{-\kappa^2 T^2/2}$ \cite{WanganddeLangePRB2012}. Let us also assume that we know $b$ perfectly. That is, we are not sure if there is a magnetic field or not, but if there is a magnetic field present, we know that it is $B(t) = b$. In this case, $\mu = e^{-i2\pi b\gamma T}$, which leads to, assuming $P_1 = P_0 = 1/2$,
\begin{equation}
P_e = \frac{1}{2}\left[ 1 - e^{-\kappa^2 T^2/2} |\sin(\pi \gamma b T)|\right].
\end{equation}
Our objective then is to find $T$ such that $P_e$ is minimized. Intuitively, this makes sense because for small $T$, the phase difference is too small to discriminate the states, while for large $T$, decoherence is too dominant. Therefore, an optimal value of $T$ needs to be chosen so as to achieve the minimum possible $P_e$. The corresponding $\mu$ can then be calculated, leading to $\chi$, and thus $\Pi_0$ and $\Pi_1$. This task is performed in Fig.~\ref{firstfigure}, where we show the detection of magnetic fields of various strengths using noise parameters drawn from recently performed experiments \cite{deLangePRL2011, WanganddeLangePRB2012}. It is clear from Fig.~\ref{DCfielddetection} that, unfortunately, it is challenging to detect DC magnetic fields of strength $1 \,\mu\text{T}$ and below using NV centers. However, detection of stronger magnetic fields such as those of strength $50 \,\mu\text{T}$ and above is very much possible. It is also clear from Fig.~\ref{DCmeasurement} that the optimum measurement depends on the magnetic field being detected and is not fixed. This should be contrasted with the situation for the measurement of magnetic fields (see, for instance, Ref.~\cite{deLangePRL2011}) where measurements are performed in the basis $\lbrace \ket{+},\ket{-}\rbrace$ with $\ket{\pm} = \frac{1}{\sqrt{2}}(\ket{0} \pm i\ket{1})$.

\begin{figure}[b]
\centering
\mbox{\hspace{-3mm}\subfigure{\includegraphics[scale = 0.4]{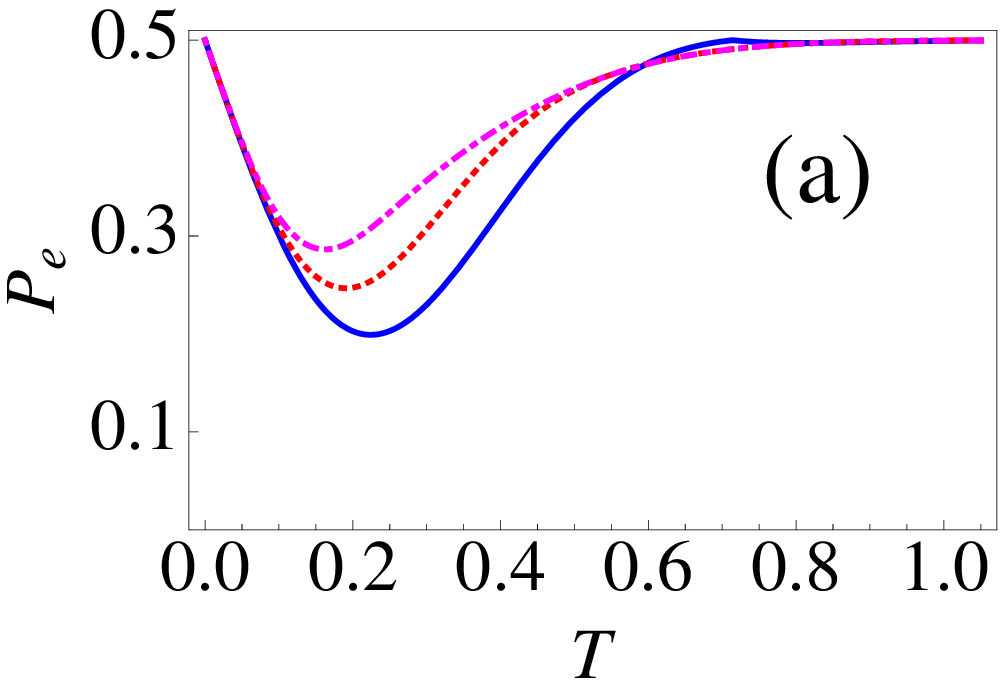}\label{DCfieldunknownmagnitude} }\hspace{-2mm} 
\subfigure{\includegraphics[scale = 0.4]{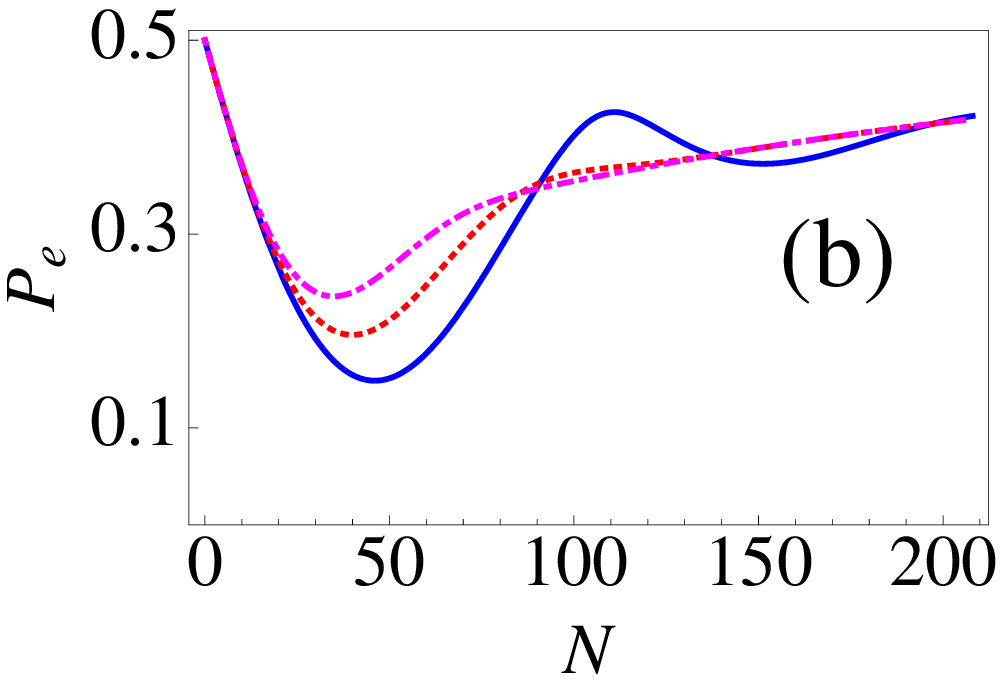}\label{FielddetectionCP} }}
\caption{(color online) (a) Error probability $P_e$ for a DC field as a function of total time $T$ (in $\mu$s) for $\sigma_b = 1\, \mu\text{T}$ (solid, blue line), $\sigma_b = 25 \,\mu\text{T}$ (dotted, red line) and $\sigma_b = 50 \,\mu\text{T}$ (dot-dashed, magenta line). Here we have used $b_0 = 50 \,\mu\text{T}$. (b) Error probability $P_e$ for an AC field as a function of the number of applied pulses $N$ for $\sigma_b = 0.2\, \mu\text{T}$ (solid, blue line), $\sigma_b = 0.4 \,\mu \text{T}$ (dotted, red line) and $\sigma_b = 0.6 \,\mu\text{T}$ (dot-dashed, magenta line). Here, $b_0 = 1 \,\mu\text{T}$ and $f = 1$ MHz.}
\label{secondfigure}
\end{figure}

A more realistic scenario is to consider the situation where we do not know $b$ precisely. Rather, $b$ is known with some probability distribution that we assume to be $P(b) = \frac{1}{\sqrt{2\pi \sigma_b^2}} e^{-(b - b_0)^2/2\sigma_b^2}.$
In other words, we suspect that there is a magnetic field of strength around $b_0$ and we are trying to find if there is indeed such a magnetic field present or not. In this case,  
$$ \mu = \int P(b) e^{-i2\pi \gamma b T} db = e^{-i2\pi \gamma b_0 T} e^{-2\pi^2 \gamma^2 T^2 \sigma_b^2}, $$
while $\nu$ remains the same. Once again, we find $T$ that minimizes $P_e$, then the corresponding $\mu$ and thereby $\Pi_0$ and $\Pi_1$. A plot of $P_e$ against $T$ is shown in Fig.~\ref{DCfieldunknownmagnitude}. Comparing with Fig.~\ref{DCfielddetection}, one can see that the less precisely we know the magnetic field, the less reliably we can say whether or not a magnetic field is present.

\begin{figure}[t]
\centering
{\includegraphics[scale = 0.5]{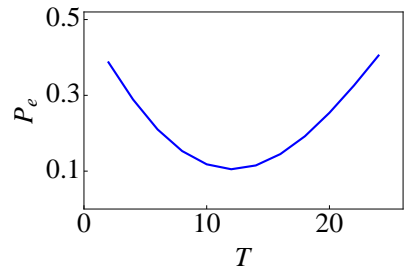}}
\caption{(color online) Error probability $P_e$ as a function of time $T$ (in $\mu$s) for the bichromatic magnetic field $B(t) = b_1 \sin(2\pi f_1 t) + b_2 \sin(2\pi f_2 t)$ with $b_1 = 1 \, \mu\text{T}$, $b_2 = 2 \, \mu\text{T}$, $f_1 = 1$ MHz and  $f_2 = 1.5$ MHz. Here we have assumed, for simplicity, that the magnetic field parameters are known perfectly.  }
\label{bichromatic}
\end{figure}

\textit{AC field detection.} We have seen that using a NV center, it is difficult to detect static magnetic fields of strength $1 \, \mu\text{T}$ and below. This is not surprising - DC magnetic fields of a similar magnitude cannot be reliably measured using NV centers either. The problem is decoherence. To eliminate decoherence, one possible solution is to use dynamical decoupling techniques \cite{ViolaPRA1998,LloydPRL1999,BiercukNature2009,LiuNature2009,HansonScience2010,RyanPRL2010,NaydenovPRB2011,
WanganddeLangePRB2012,BarGillNatCommun2012,ZhaoPRB2012,WitzelPRB2012}, whereby a sequence of rapid control pulses are applied to the NV center. These pulses effectively keep on flipping the sign of the NV center-spin bath interaction Hamiltonian. Thus, if these pulses are applied rapidly enough, the effect of the spin bath is largely eliminated. Unfortunately, for a DC field, the effect of the static field that we are trying to detect is also eliminated. But for an AC field, the direction of the magnetic field also keeps on changing direction. Dynamical decoupling can then still be used, provided that we apply the pulses at (or near) the nodes of the magnetic field so that the phase difference between the NV center states keeps on accumulating. Let us examine this more quantitatively. For $B(t) = b \cos(2\pi f t)$, we apply the Carr-Purcell-Meiboom-Gill (CPMG) pulse sequence, specified by $[U(\tau/2)R(\pi)U(\tau)R(\pi)U(\tau/2)]^{N/2}$. This means that we allow the NV center to evolve freely for time $\tau/2$ (denoted by $U(\tau/2)$), then we apply a control pulse specified by $R(\pi) = e^{-i\pi \sigma_x/2}$, followed by $U(\tau)$, then another control pulse, and finally we have $U(\tau/2)$ again. This whole cycle is then repeated $N/2$ times, where $N$ is the number of pulses applied. To ensure that we apply the pulses at the nodes of the magnetic field, we choose $\tau = 1/2f$. Once again assuming that $P(b) = \frac{1}{\sqrt{2\pi \sigma_b^2}} e^{-(b - b_0)^2/2\sigma_b^2}$, we find $\mu = e^{-i2N\gamma b_0/f} e^{-2N^2 \gamma^2 \sigma_b^2/f^2}$. The computation for $\nu$, on the other hand, is considerably more involved. In the presence of the pulses, $\nu = \langle \exp(-i\int_0^T \xi(t) B_d(t)\, dt )\rangle$, where $\xi(t)$, which can assume values $+1$ or $-1$, takes into account the effect of the pulses by switching sign whenever a pulse is applied. It can then be shown that $\nu = e^{-\kappa^2 W(T)}$, with $W(T) = \int_0^T e^{-Rs} p(s)\,ds$, $R = 1/\tau_c$, and $p(s) = \int_0^{T-s}\xi(t)\xi(t+s)\,dt$ (see Refs.~\cite{WanganddeLangePRB2012,ChaudhryPRA2014NV} for details). Thus, $\nu$ for any pulse sequence can be calculated, at least numerically. For the CPMG sequence, it is relatively straightforward, though laborious, to derive the analytic form for $\nu$ which can be found in the Supplementary Material \cite{supple}. Using the expressions for $\nu$ and $\mu$, we find $N$ (or, in other words, $T = N\tau$) so as to obtain the smallest possible value of $P_e$. This is illustrated in Fig.~\ref{FielddetectionCP}. The key point is that now $\nu$ is much closer to $1$, leading to considerably smaller values of $P_e$ for a field strength of around $1 \, \mu\text{T}$. Moreover, it is possible that we also do not know, for instance, the frequency of the magnetic field precisely. In this case, in the computation for $\mu$, we can include a probability distribution for frequency just like we did for the amplitude of the magnetic field.

One can also consider other kinds of oscillating magnetic fields. 
For example, consider $B(t) = b_1 \sin(2\pi f_1 t) + b_2 \sin(2\pi f_2 t)$, with $f_1 \neq f_2$. The control pulses are again applied at the nodes of the magnetic field, and $\mu = e^{-i2\pi \gamma \int_0^T |B(t')| dt'}$ (assuming that the field is perfectly known) and $\nu$ can be determined numerically. Results are illustrated in Fig.~\ref{bichromatic}, showing that by suitably choosing $T$ we can detect more complicated fields reliably as well.

\begin{figure}[t]
\centering
{\includegraphics[scale = 0.5]{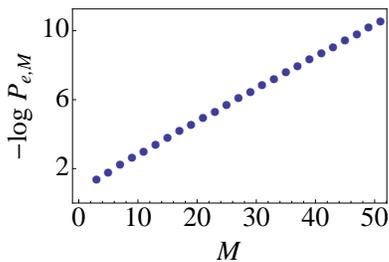}}
\caption{(color online) $-\log P_{e,M}$ as a function the number of number of measurements, where we try to detect the magnetic field $B(t) = b_0 \cos(2\pi f t)$ using the CPMG sequence. The parameters are same as used in Fig.~\ref{FielddetectionCP}, except that now $\sigma_b \rightarrow 0$. In this case, $P_{e,1} \approx 0.124$.}
\label{Multiplemeasurements}
\end{figure}

\textit{Using multiple copies.} The error probabilities we have found until now utilize only one copy of the NV center. In practice, multiple copies of the NV center, which are also used when we are measuring a magnetic field, can be used to reduce the error probability $P_e$ even further \cite{HigginsPRA2011}. Let us define $C_{0|1} = \text{Tr}[\rho_1 \Pi_0]$ and $C_{1|0} = \text{Tr}[\rho_0 \Pi_1]$ as the conditional error probabilities. Then, with $M$ fixed local measurements on $M$ copies, the total error probability is 
\begin{align}
P_{e,M} &= P_1 \sum_{m = 0}^{\floor{M/2}} \binom{M}{m} (1 - C_{0|1})^m C_{0|1}^m \, + \notag \\
&P_0 \sum_{m = 0}^{\floor{M/2}} \binom{M}{m} (1 - C_{1|0})^m C_{1|0}^m,
\end{align}
for $M$ odd, while if $M$ is even, then $P_{e,M} = P_{e,M-1}$. Fig.~\ref{Multiplemeasurements} shows clearly how, by using multiple NV centers (or using the same NV center repeatedly), we can very reliably detect the magnetic field. It should also be noted that we are performing the same measurement again and again. Even better results could be obtained if we keep on updating the measurement that should be performed \cite{HigginsPRA2011}, which is the subject of a future work.

\begin{figure}[t]
\centering
{\includegraphics[scale = 0.5]{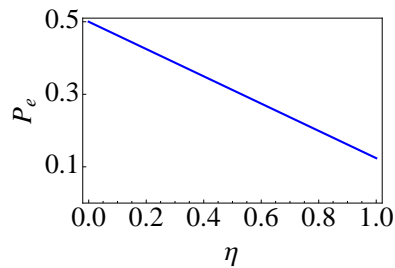}}
\caption{(color online) Error probability $P_e$ as a function of the photon detection efficiency $\eta$, where again we  try to detect the magnetic field $B(t) = b_0 \cos(2\pi f t)$ using the CPMG sequence. Once again, the parameters are same as used in Fig.~\ref{FielddetectionCP}, except that now $\sigma_b \rightarrow 0$.}
\label{finiteefficiency}
\end{figure}

\textit{Imperfect photon detection efficiency.} Finally, before concluding, we explain how the optimum measurement (specified by $\chi$) can be performed. Measurement in the basis $\lbrace\ket{\phi_+},\ket{\phi_-}\rbrace$ cannot be performed directly. Rather, we can use spin-dependent fluorescence to perform measurements in the basis $\lbrace\ket{0},\ket{1}\rbrace$ \cite{RondinRepProgPhys2014}. Therefore, in order to effectively measure in the $\lbrace\ket{\phi_+},\ket{\phi_-}\rbrace$ basis, we need to first apply a rotation operator $U_R$ to the NV center state. Labeling the $\sigma_z$ eigenvalues as $n$, we find that $p(n = 1) = \text{Tr}[U_R^\dagger \rho U_R \ket{0}\bra{0}] = \text{Tr}[\rho \ket{\phi_+}\bra{\phi_+}]$ and $p(n = -1) = \text{Tr}[U_R^\dagger \rho U_R \ket{1}\bra{1}] = \text{Tr}[\rho \ket{\phi_-}\bra{\phi_-}]$, where $\rho$ is the NV center state. For the optimum measurement found before,  $U_R = e^{-i\chi \sigma_z/2} e^{i\pi \sigma_y/4}$. However, this measurement is not perfect since the fluorescent photons cannot be captured with $100\%$ efficiency, and this is a major source of error in NV center magnetometry \cite{RondinRepProgPhys2014}. In order to model the imperfect detection, we introduce the conditional probabilities $p(m = 1|n = 1) = \eta$, $p(m = 0|n = 1) = 1 - \eta$, $p(m = 1|n = -1) = 0$ and $p(m = 0|n = 1) = 1$, where $m = 0$ means no photon detected, while $m = 1$ means at least one photon detected. In particular, $\eta = 1$ implies perfect measurement, while $\eta = 0$ implies a totally imperfect measurement. By using this simple model, we are able to deduce (see the Supplementary Material \cite{supple} for details) that 
\begin{equation}
P_e = P_1 - \frac{\eta}{2}\left[(P_1 - P_0) - \nu(P_1 \text{Re}(\mu e^{i\chi}) - P_0 \cos \chi)\right].
\end{equation}
Moreover, we can also show that $\eta \neq 1$ does not change the measurement that should be performed \cite{supple}. In Fig.~\ref{finiteefficiency}, we have plotted the error probability as a function of $\eta$, showing that we do not need perfect measurement to detect magnetic fields. Of course, by using multiple copies of the NV center, this error probability can be reduced further.

\textit{Conclusion.} To conclude, we have considered the detection of weak magnetic fields using NV centers. We have shown that oscillating magnetic fields of strength in the microtesla regime can be easily detected, especially if multiple NV centers are utilized. Besides constructing the optimum measurement that should be performed, we also took into account the imperfect measurement of the NV center state. This work should open up new directions in the use of NV centers as magnetic sensors.

This work is supported by the Singapore National Research Foundation under NRF Grant No. NRF-NRFF2011-07 and the LUMS Startup Grant. Discussions with M.~Tsang and R.~Nair are gratefully acknowledged.

\pagebreak

\onecolumngrid

\setcounter{equation}{0}

\section*{Supplementary Material: Detecting magnetic fields using Nitrogen-Vacancy Centers}

In this Supplementary material, we use the same notation as introduced in our main text.

\section{Calculating dephasing with the CPMG sequence}

As explained in the main text, we suppose that the NV center experiences a classical Gaussian noise field $B_d(t)$ with zero mean and correlation function 
\begin{equation}
\langle B_d(0) B_d(t) \rangle = \kappa^2 e^{-|t|/\tau_c},
\end{equation}  
where $\tau_c$ is the correlation time and $\kappa$ describes the coupling between the NV center and the P1 centers. With the applied pulses, we can write 
\begin{equation*}
S(T) = \left\langle \exp \left( -i\int_0^T \xi(t) B_d(t) \, dt \right) \right\rangle,
\end{equation*}
where $\xi(t)$, which can assume the values $+1$ or $-1$, takes into account the effect of the pulses by switching sign whenever a pulse is applied. It can then be shown that (see Ref.~\cite{suppleWanganddeLangePRB2012,suppleChaudhryPRA2014NV} for a detailed derivation)
\begin{equation}
S(T) = \exp \left[ -\kappa^2 W(T) \right],
\end{equation}
where $W(T) = \int_0^T e^{-Rs} p(s) \, ds$, with $R = 1/\tau_c$, and $p(s) = \int_0^{T - s} \xi(t) \xi(t + s)\, dt$ depends on the pulse sequence applied. Using this formalism, $W(T)$ can be evaluated for different pulse sequences. For the CPMG sequence, we can write \cite{suppleWanganddeLangePRB2012,suppleChaudhryPRA2014NV}
\begin{equation}
W_{\text{CP}}(T) = \Gamma_N(Q_{11}^{\text{CP}} + Q_{12}^{\text{CP}}) - P_N Q_{12}^{\text{CP}},
\end{equation}
with 
\begin{align*}
P_N &= \frac{1 - e^{-N\delta}}{1 - e^{-2\delta}}, \notag \\
\Gamma_N &= \frac{0.5N - (0.5N + 1)e^{-2\delta} + e^{-(N + 2)\delta}}{(1 - e^{-2\delta})^2}, \notag \\
Q_{11}^{\text{CP}} &= \frac{1}{R^2} \left[ 2\delta - 5 + 4(e^{-\frac{\delta}{2}} + e^{-\delta} - e^{-\frac{3\delta}{2}}) + e^{-2\delta}\right], \\
Q_{12}^{\text{CP}} &= \frac{1}{R^2} \left[ 1 - 4(e^{-\frac{\delta}{2}} - e^{-\delta} - e^{-\frac{3\delta}{2}}) - (2\delta + 5)e^{-2\delta}\right],
\end{align*}
and $\delta = R\tau$.

\section{Dealing with imperfect photon detection efficiency}

As stated in the main text, we can only perform measurements in the basis $\lbrace \ket{0},\ket{1} \rbrace$ via spin-dependent fluorescence. Let us briefly explain here what this means. The negatively charged NV center forms a spin triplet system in its ground state. In the presence of a magnetic field, the degeneracy of the $m_S = \pm 1$ levels is lifted. The $m_S = 0$ and $m_S = -1$ form an effective two-level system. This is precisely the two-level system that has been used in this work, with the identification $\ket{0} \rightarrow \ket{m_S = 0}$ and $\ket{1} \rightarrow \ket{m_S = -1}$. Now, the energy level $\ket{m_S = 0} = \ket{0}$ shows far more fluorescence  than the $\ket{m_S = -1} = \ket{1}$ level, and this is the property that is used to perform measurements on the NV center in the $\lbrace \ket{0}, \ket{1} \rbrace$ basis. If a fluorescent photon is detected, we deduce the state to be $\ket{0}$, otherwise the state is $\ket{1}$. However, this measurement is not perfect due to limited photon detection efficiency - some of the fluorescent photons are lost, while some are not detected due to imperfect detectors. This imperfect measurement is a major source of errors in NV center magnetometry \cite{suppleRondin2014}, and this is what we now analyze. 

To start, recall that the error probability is given by $P_e = P_0 \text{Tr} (\rho_0 \Pi_1) + P_1 \text{Tr} (\rho_1 \Pi_0)$, where $\Pi_0$ and $\Pi_1$ are orthogonal projectors in our two-dimensional Hilbert space. Since we can only perform measurements in the $\lbrace \ket{0}, \ket{1} \rbrace$ basis, we define the unitary operator $U_R$ as $U_R \ket{1}\bra{1} U_R^\dagger = \Pi_0$ and $U_R \ket{0}\bra{0} U_R^\dagger = \Pi_1$. Note that we know that $U_R = e^{-i\chi \sigma_z/2} e^{i\pi \sigma_y/4}$ for the case of ideal measurement, but here we allow for the possibility that $U_R$ may change due to imperfect photon detection. It follows that $P_e = P_0 \text{Tr} (U_R^\dagger \rho_0 U_R \ket{0}\bra{0}) + P_1 \text{Tr} (U_R^\dagger \rho_1 U_R \ket{1}\bra{1})$, showing explicitly how to perform measurements in the basis $\lbrace \ket{0},\ket{1} \rbrace$ by first performing the unitary operator $U_R$ on the NV center state. This unitary operator can be implemented experimentally \cite{suppleKennedyNV}.  

Now, the way in which we have written the error probability, namely $P_e = P_0 \text{Tr} (U_R^\dagger \rho_0 U_R \ket{0}\bra{0}) + P_1 \text{Tr} (U_R^\dagger \rho_1 U_R \ket{1}\bra{1})$, means that if we detect a fluorescent photon, then we deduce that there is a magnetic field, while we say that there is no magnetic field otherwise. As already explained, this photon detection is not perfect. In order to model the detection error, we introduce the conditional probabilities $p(m = 1|n = 1) = \eta$, $p(m = 0|n = 1) = 1 - \eta$, $p(m = 1|n = -1) = 0$ and $p(m = 0|n = 1) = 1$, where $m = 0$ means no photon detected, while $m = 1$ means at least one photon detected. For the state $\ket{0}$, $n$ is $+1$, while $n$ is $-1$ for the state $\ket{1}$. It then follows that 
\begin{align*}
p(m = 1) &= \eta p(n = 1), \\
p(m = 0) &= (1 - \eta)p(n = 1) + p(n = -1), 
\end{align*}
where $p(n = 1) = \text{Tr}[U_R^\dagger \rho U_R \ket{0}\bra{0}]$ and $p(n = -1) = \text{Tr}[U_R^\dagger \rho U_R \ket{1}\bra{1}]$ (here $\rho$ refers to either $\rho_0$ or $\rho_1$). It then follows that 
\begin{equation*}
P_e = P_0 \eta \opav{0}{U_R^\dagger \rho_0 U_R}{0} + P_1 [(1 - \eta)\opav{0}{U_R^\dagger \rho_1 U_R}{0} + \opav{1}{U_R^\dagger \rho_1 U_R}{1}].
\end{equation*}
Using the fact that $\ket{0}\bra{0} + \ket{1}\bra{1} = 1$, we can simplify this to 
\begin{equation}
\label{minerror}
P_e = P_1 + \eta[P_0 \opav{0}{U_R^\dagger \rho_0 U_R}{0} - P_1 \opav{0}{U_R^\dagger \rho_1 U_R}{0}].
\end{equation}
This means that to obtain the minimum possible value of the error probability $P_e$, we should choose $U_R$ such that the expression in the square brackets is minimized. This minimization is obviously independent of $\eta$, and we have already performed this minimization for $\eta = 1$. Thus, even if $\eta \neq 1$, we should still use $U_R = e^{-i\chi \sigma_z/2} e^{i\pi \sigma_y/4}$. Substituting this optimal $U_R$ in \eqref{minerror}, we find that 
\begin{equation}
P_e = P_1 - \frac{\eta}{2}\left[(P_1 - P_0) - \nu(P_1 \text{Re}(\mu e^{i\chi}) - P_0 \cos \chi)\right].
\end{equation}


\begin{thebibliography}{99}%
\makeatletter
\providecommand \@ifxundefined [1]{%
 \@ifx{#1\undefined}
}%
\providecommand \@ifnum [1]{%
 \ifnum #1\expandafter \@firstoftwo
 \else \expandafter \@secondoftwo
 \fi
}%
\providecommand \@ifx [1]{%
 \ifx #1\expandafter \@firstoftwo
 \else \expandafter \@secondoftwo
 \fi
}%
\providecommand \natexlab [1]{#1}%
\providecommand \enquote  [1]{``#1''}%
\providecommand \bibnamefont  [1]{#1}%
\providecommand \bibfnamefont [1]{#1}%
\providecommand \citenamefont [1]{#1}%
\providecommand \href@noop [0]{\@secondoftwo}%
\providecommand \href [0]{\begingroup \@sanitize@url \@href}%
\providecommand \@href[1]{\@@startlink{#1}\@@href}%
\providecommand \@@href[1]{\endgroup#1\@@endlink}%
\providecommand \@sanitize@url [0]{\catcode `\\12\catcode `\$12\catcode
  `\&12\catcode `\#12\catcode `\^12\catcode `\_12\catcode `\%12\relax}%
\providecommand \@@startlink[1]{}%
\providecommand \@@endlink[0]{}%
\providecommand \url  [0]{\begingroup\@sanitize@url \@url }%
\providecommand \@url [1]{\endgroup\@href {#1}{\urlprefix }}%
\providecommand \urlprefix  [0]{URL }%
\providecommand \Eprint [0]{\href }%
\providecommand \doibase [0]{http://dx.doi.org/}%
\providecommand \selectlanguage [0]{\@gobble}%
\providecommand \bibinfo  [0]{\@secondoftwo}%
\providecommand \bibfield  [0]{\@secondoftwo}%
\providecommand \translation [1]{[#1]}%
\providecommand \BibitemOpen [0]{}%
\providecommand \bibitemStop [0]{}%
\providecommand \bibitemNoStop [0]{.\EOS\space}%
\providecommand \EOS [0]{\spacefactor3000\relax}%
\providecommand \BibitemShut  [1]{\csname bibitem#1\endcsname}%
\let\auto@bib@innerbib\@empty
\bibitem [{\citenamefont {Freeman}\ and\ \citenamefont
  {Choi}(2001)}]{FreemanScience2001}%
  \BibitemOpen
  \bibfield  {author} {\bibinfo {author} {\bibfnamefont {M.~R.}\ \bibnamefont
  {Freeman}}\ and\ \bibinfo {author} {\bibfnamefont {B.~C.}\ \bibnamefont
  {Choi}},\ }\href@noop {} {\bibfield  {journal} {\bibinfo  {journal}
  {Science}\ }\textbf {\bibinfo {volume} {294}},\ \bibinfo {pages} {1484}
  (\bibinfo {year} {2001})}\BibitemShut {NoStop}%
\bibitem [{\citenamefont {Budker}\ and\ \citenamefont
  {Romalis}(2007)}]{Budkermagnetometry}%
  \BibitemOpen
  \bibfield  {author} {\bibinfo {author} {\bibfnamefont {D.}~\bibnamefont
  {Budker}}\ and\ \bibinfo {author} {\bibfnamefont {M.}~\bibnamefont
  {Romalis}},\ }\href {\doibase 10.1038/nphys566} {\bibfield  {journal}
  {\bibinfo  {journal} {Nat. Phys.}\ }\textbf {\bibinfo {volume} {3}},\
  \bibinfo {pages} {227} (\bibinfo {year} {2007})}\BibitemShut {NoStop}%
\bibitem [{\citenamefont {Greenberg}(1998)}]{GreenbergRevModPhys1998}%
  \BibitemOpen
  \bibfield  {author} {\bibinfo {author} {\bibfnamefont {Y.~S.}\ \bibnamefont
  {Greenberg}},\ }\href {\doibase 10.1103/RevModPhys.70.175} {\bibfield
  {journal} {\bibinfo  {journal} {Rev. Mod. Phys.}\ }\textbf {\bibinfo {volume}
  {70}},\ \bibinfo {pages} {175} (\bibinfo {year} {1998})}\BibitemShut
  {NoStop}%
\bibitem [{\citenamefont {Grinolds}\ \emph {et~al.}(2011)\citenamefont
  {Grinolds}, \citenamefont {Maletinsky}, \citenamefont {Hong}, \citenamefont
  {Lukin}, \citenamefont {Walsworth},\ and\ \citenamefont
  {Yacoby}}]{GrinoldsNatPhys2011}%
  \BibitemOpen
  \bibfield  {author} {\bibinfo {author} {\bibfnamefont {M.~S.}\ \bibnamefont
  {Grinolds}}, \bibinfo {author} {\bibfnamefont {P.}~\bibnamefont
  {Maletinsky}}, \bibinfo {author} {\bibfnamefont {S.}~\bibnamefont {Hong}},
  \bibinfo {author} {\bibfnamefont {M.~D.}\ \bibnamefont {Lukin}}, \bibinfo
  {author} {\bibfnamefont {R.~L.}\ \bibnamefont {Walsworth}}, \ and\ \bibinfo
  {author} {\bibfnamefont {A.}~\bibnamefont {Yacoby}},\ }\href
  {http://dx.doi.org/10.1038/nphys1999} {\bibfield  {journal} {\bibinfo
  {journal} {Nat. Phys.}\ }\textbf {\bibinfo {volume} {7}},\ \bibinfo {pages}
  {687} (\bibinfo {year} {2011})}\BibitemShut {NoStop}%
\bibitem [{\citenamefont {Jelezko}\ \emph {et~al.}(2002)\citenamefont
  {Jelezko}, \citenamefont {Popa}, \citenamefont {Gruber}, \citenamefont
  {Tietz}, \citenamefont {Wrachtrup}, \citenamefont {Nizovtsev},\ and\
  \citenamefont {Kilin}}]{JelezkoApplPhys2002}%
  \BibitemOpen
  \bibfield  {author} {\bibinfo {author} {\bibfnamefont {F.}~\bibnamefont
  {Jelezko}}, \bibinfo {author} {\bibfnamefont {I.}~\bibnamefont {Popa}},
  \bibinfo {author} {\bibfnamefont {A.}~\bibnamefont {Gruber}}, \bibinfo
  {author} {\bibfnamefont {C.}~\bibnamefont {Tietz}}, \bibinfo {author}
  {\bibfnamefont {J.}~\bibnamefont {Wrachtrup}}, \bibinfo {author}
  {\bibfnamefont {A.}~\bibnamefont {Nizovtsev}}, \ and\ \bibinfo {author}
  {\bibfnamefont {S.}~\bibnamefont {Kilin}},\ }\href {\doibase
  http://dx.doi.org/10.1063/1.1507838} {\bibfield  {journal} {\bibinfo
  {journal} {Appl. Phys. Lett.}\ }\textbf {\bibinfo {volume} {81}},\ \bibinfo
  {pages} {2160} (\bibinfo {year} {2002})}\BibitemShut {NoStop}%
\bibitem [{\citenamefont {Doherty}\ \emph {et~al.}(2013)\citenamefont
  {Doherty}, \citenamefont {Manson}, \citenamefont {Delaney}, \citenamefont
  {Jelezko}, \citenamefont {Wrachtrup},\ and\ \citenamefont
  {Hollenberg}}]{DohertyPhysRep2013}%
  \BibitemOpen
  \bibfield  {author} {\bibinfo {author} {\bibfnamefont {M.~W.}\ \bibnamefont
  {Doherty}}, \bibinfo {author} {\bibfnamefont {N.~B.}\ \bibnamefont {Manson}},
  \bibinfo {author} {\bibfnamefont {P.}~\bibnamefont {Delaney}}, \bibinfo
  {author} {\bibfnamefont {F.}~\bibnamefont {Jelezko}}, \bibinfo {author}
  {\bibfnamefont {J.}~\bibnamefont {Wrachtrup}}, \ and\ \bibinfo {author}
  {\bibfnamefont {L.~C.~L.}\ \bibnamefont {Hollenberg}},\ }\href@noop {}
  {\bibfield  {journal} {\bibinfo  {journal} {Phys.~Rep.}\ }\textbf {\bibinfo
  {volume} {528}},\ \bibinfo {pages} {1} (\bibinfo {year} {2013})}\BibitemShut
  {NoStop}%
\bibitem [{\citenamefont {Taylor}\ \emph {et~al.}(2008)\citenamefont {Taylor},
  \citenamefont {Cappellaro}, \citenamefont {Childress}, \citenamefont {Jiang},
  \citenamefont {Budker}, \citenamefont {Hemmer}, \citenamefont {Yacoby},
  \citenamefont {Walsworth},\ and\ \citenamefont {Lukin}}]{TaylorNatPhys2008}%
  \BibitemOpen
  \bibfield  {author} {\bibinfo {author} {\bibfnamefont {J.~M.}\ \bibnamefont
  {Taylor}}, \bibinfo {author} {\bibfnamefont {P.}~\bibnamefont {Cappellaro}},
  \bibinfo {author} {\bibfnamefont {L.}~\bibnamefont {Childress}}, \bibinfo
  {author} {\bibfnamefont {L.}~\bibnamefont {Jiang}}, \bibinfo {author}
  {\bibfnamefont {D.}~\bibnamefont {Budker}}, \bibinfo {author} {\bibfnamefont
  {P.~R.}\ \bibnamefont {Hemmer}}, \bibinfo {author} {\bibfnamefont
  {A.}~\bibnamefont {Yacoby}}, \bibinfo {author} {\bibfnamefont
  {R.}~\bibnamefont {Walsworth}}, \ and\ \bibinfo {author} {\bibfnamefont
  {M.~D.}\ \bibnamefont {Lukin}},\ }\href@noop {} {\bibfield  {journal}
  {\bibinfo  {journal} {Nat.~Phys.}\ }\textbf {\bibinfo {volume} {4}},\
  \bibinfo {pages} {810} (\bibinfo {year} {2008})}\BibitemShut {NoStop}%
\bibitem [{\citenamefont {Maze}\ \emph {et~al.}(2008)\citenamefont {Maze},
  \citenamefont {Stanwix}, \citenamefont {Hodges}, \citenamefont {Hong},
  \citenamefont {Taylor}, \citenamefont {Cappellaro}, \citenamefont {Jiang},
  \citenamefont {Dutt}, \citenamefont {Togan}, \citenamefont {Zibrov},
  \citenamefont {Yacoby}, \citenamefont {Walsworth},\ and\ \citenamefont
  {Lukin}}]{MazeNature2008}%
  \BibitemOpen
  \bibfield  {author} {\bibinfo {author} {\bibfnamefont {J.~R.}\ \bibnamefont
  {Maze}}, \bibinfo {author} {\bibfnamefont {P.~L.}\ \bibnamefont {Stanwix}},
  \bibinfo {author} {\bibfnamefont {J.~S.}\ \bibnamefont {Hodges}}, \bibinfo
  {author} {\bibfnamefont {S.}~\bibnamefont {Hong}}, \bibinfo {author}
  {\bibfnamefont {J.~M.}\ \bibnamefont {Taylor}}, \bibinfo {author}
  {\bibfnamefont {P.}~\bibnamefont {Cappellaro}}, \bibinfo {author}
  {\bibfnamefont {L.}~\bibnamefont {Jiang}}, \bibinfo {author} {\bibfnamefont
  {M.~V.~G.}\ \bibnamefont {Dutt}}, \bibinfo {author} {\bibfnamefont
  {E.}~\bibnamefont {Togan}}, \bibinfo {author} {\bibfnamefont {A.~S.}\
  \bibnamefont {Zibrov}}, \bibinfo {author} {\bibfnamefont {A.}~\bibnamefont
  {Yacoby}}, \bibinfo {author} {\bibfnamefont {R.~L.}\ \bibnamefont
  {Walsworth}}, \ and\ \bibinfo {author} {\bibfnamefont {M.~D.}\ \bibnamefont
  {Lukin}},\ }\href@noop {} {\bibfield  {journal} {\bibinfo  {journal}
  {Nature}\ }\textbf {\bibinfo {volume} {455}},\ \bibinfo {pages} {644}
  (\bibinfo {year} {2008})}\BibitemShut {NoStop}%
\bibitem [{\citenamefont {Balasubramanian}\ \emph {et~al.}(2008)\citenamefont
  {Balasubramanian}, \citenamefont {Chan}, \citenamefont {Kolesov},
  \citenamefont {Al-Hmoud}, \citenamefont {Tisler}, \citenamefont {Shin},
  \citenamefont {Kim}, \citenamefont {Wojcik}, \citenamefont {Hemmer},
  \citenamefont {Krueger}, \citenamefont {Hanke}, \citenamefont
  {Leitenstorfer}, \citenamefont {Bratschitsch}, \citenamefont {Jelezko},\ and\
  \citenamefont {Wrachtrup}}]{BalasubramanianNature2008}%
  \BibitemOpen
  \bibfield  {author} {\bibinfo {author} {\bibfnamefont {G.}~\bibnamefont
  {Balasubramanian}}, \bibinfo {author} {\bibfnamefont {I.~Y.}\ \bibnamefont
  {Chan}}, \bibinfo {author} {\bibfnamefont {R.}~\bibnamefont {Kolesov}},
  \bibinfo {author} {\bibfnamefont {M.}~\bibnamefont {Al-Hmoud}}, \bibinfo
  {author} {\bibfnamefont {J.}~\bibnamefont {Tisler}}, \bibinfo {author}
  {\bibfnamefont {C.}~\bibnamefont {Shin}}, \bibinfo {author} {\bibfnamefont
  {C.}~\bibnamefont {Kim}}, \bibinfo {author} {\bibfnamefont {A.}~\bibnamefont
  {Wojcik}}, \bibinfo {author} {\bibfnamefont {P.~R.}\ \bibnamefont {Hemmer}},
  \bibinfo {author} {\bibfnamefont {A.}~\bibnamefont {Krueger}}, \bibinfo
  {author} {\bibfnamefont {T.}~\bibnamefont {Hanke}}, \bibinfo {author}
  {\bibfnamefont {A.}~\bibnamefont {Leitenstorfer}}, \bibinfo {author}
  {\bibfnamefont {R.}~\bibnamefont {Bratschitsch}}, \bibinfo {author}
  {\bibfnamefont {F.}~\bibnamefont {Jelezko}}, \ and\ \bibinfo {author}
  {\bibfnamefont {J.}~\bibnamefont {Wrachtrup}},\ }\href@noop {} {\bibfield
  {journal} {\bibinfo  {journal} {Nature}\ }\textbf {\bibinfo {volume} {455}},\
  \bibinfo {pages} {648} (\bibinfo {year} {2008})}\BibitemShut {NoStop}%
\bibitem [{\citenamefont {Chang}\ \emph {et~al.}(2008)\citenamefont {Chang},
  \citenamefont {Lee}, \citenamefont {Chen}, \citenamefont {Chang},
  \citenamefont {Tsai}, \citenamefont {Fu}, \citenamefont {Lim}, \citenamefont
  {Tzeng}, \citenamefont {Fang}, \citenamefont {Han}, \citenamefont {Chang},\
  and\ \citenamefont {Fann}}]{ChangNatNano2008}%
  \BibitemOpen
  \bibfield  {author} {\bibinfo {author} {\bibfnamefont {Y.-R.}\ \bibnamefont
  {Chang}}, \bibinfo {author} {\bibfnamefont {H.-Y.}\ \bibnamefont {Lee}},
  \bibinfo {author} {\bibfnamefont {K.}~\bibnamefont {Chen}}, \bibinfo {author}
  {\bibfnamefont {C.-C.}\ \bibnamefont {Chang}}, \bibinfo {author}
  {\bibfnamefont {D.-S.}\ \bibnamefont {Tsai}}, \bibinfo {author}
  {\bibfnamefont {C.-C.}\ \bibnamefont {Fu}}, \bibinfo {author} {\bibfnamefont
  {T.-S.}\ \bibnamefont {Lim}}, \bibinfo {author} {\bibfnamefont {Y.-K.}\
  \bibnamefont {Tzeng}}, \bibinfo {author} {\bibfnamefont {C.-Y.}\ \bibnamefont
  {Fang}}, \bibinfo {author} {\bibfnamefont {C.-C.}\ \bibnamefont {Han}},
  \bibinfo {author} {\bibfnamefont {H.-C.}\ \bibnamefont {Chang}}, \ and\
  \bibinfo {author} {\bibfnamefont {W.}~\bibnamefont {Fann}},\ }\href@noop {}
  {\bibfield  {journal} {\bibinfo  {journal} {Nat.~Nanotechnol.}\ }\textbf
  {\bibinfo {volume} {3}},\ \bibinfo {pages} {284} (\bibinfo {year}
  {2008})}\BibitemShut {NoStop}%
\bibitem [{\citenamefont {Balasubramanian}\ \emph {et~al.}(2009)\citenamefont
  {Balasubramanian}, \citenamefont {Neumann}, \citenamefont {Twitchen},
  \citenamefont {Markham}, \citenamefont {Kolesov}, \citenamefont {Mizuochi},
  \citenamefont {Isoya}, \citenamefont {Achard}, \citenamefont {Beck},
  \citenamefont {Tissler}, \citenamefont {Jacques}, \citenamefont {Hemmer},
  \citenamefont {Jelezko},\ and\ \citenamefont
  {Wrachtrup}}]{BalasubramanianNatNano2009}%
  \BibitemOpen
  \bibfield  {author} {\bibinfo {author} {\bibfnamefont {G.}~\bibnamefont
  {Balasubramanian}}, \bibinfo {author} {\bibfnamefont {P.}~\bibnamefont
  {Neumann}}, \bibinfo {author} {\bibfnamefont {D.}~\bibnamefont {Twitchen}},
  \bibinfo {author} {\bibfnamefont {M.}~\bibnamefont {Markham}}, \bibinfo
  {author} {\bibfnamefont {R.}~\bibnamefont {Kolesov}}, \bibinfo {author}
  {\bibfnamefont {N.}~\bibnamefont {Mizuochi}}, \bibinfo {author}
  {\bibfnamefont {J.}~\bibnamefont {Isoya}}, \bibinfo {author} {\bibfnamefont
  {J.}~\bibnamefont {Achard}}, \bibinfo {author} {\bibfnamefont
  {J.}~\bibnamefont {Beck}}, \bibinfo {author} {\bibfnamefont {J.}~\bibnamefont
  {Tissler}}, \bibinfo {author} {\bibfnamefont {V.}~\bibnamefont {Jacques}},
  \bibinfo {author} {\bibfnamefont {P.~R.}\ \bibnamefont {Hemmer}}, \bibinfo
  {author} {\bibfnamefont {F.}~\bibnamefont {Jelezko}}, \ and\ \bibinfo
  {author} {\bibfnamefont {J.}~\bibnamefont {Wrachtrup}},\ }\href@noop {}
  {\bibfield  {journal} {\bibinfo  {journal} {Nat. Mater.}\ }\textbf {\bibinfo
  {volume} {8}},\ \bibinfo {pages} {383} (\bibinfo {year} {2009})}\BibitemShut
  {NoStop}%
\bibitem [{\citenamefont {Hall}\ \emph {et~al.}(2009)\citenamefont {Hall},
  \citenamefont {Cole}, \citenamefont {Hill},\ and\ \citenamefont
  {Hollenberg}}]{HallPRL2009}%
  \BibitemOpen
  \bibfield  {author} {\bibinfo {author} {\bibfnamefont {L.~T.}\ \bibnamefont
  {Hall}}, \bibinfo {author} {\bibfnamefont {J.~H.}\ \bibnamefont {Cole}},
  \bibinfo {author} {\bibfnamefont {C.~D.}\ \bibnamefont {Hill}}, \ and\
  \bibinfo {author} {\bibfnamefont {L.~C.~L.}\ \bibnamefont {Hollenberg}},\
  }\href {\doibase 10.1103/PhysRevLett.103.220802} {\bibfield  {journal}
  {\bibinfo  {journal} {Phys. Rev. Lett.}\ }\textbf {\bibinfo {volume} {103}},\
  \bibinfo {pages} {220802} (\bibinfo {year} {2009})}\BibitemShut {NoStop}%
\bibitem [{\citenamefont {Laraoui}\ \emph {et~al.}(2010)\citenamefont
  {Laraoui}, \citenamefont {Hodges},\ and\ \citenamefont
  {Meriles}}]{LaraouiAPL2010}%
  \BibitemOpen
  \bibfield  {author} {\bibinfo {author} {\bibfnamefont {A.}~\bibnamefont
  {Laraoui}}, \bibinfo {author} {\bibfnamefont {J.~S.}\ \bibnamefont {Hodges}},
  \ and\ \bibinfo {author} {\bibfnamefont {C.~A.}\ \bibnamefont {Meriles}},\
  }\href@noop {} {\bibfield  {journal} {\bibinfo  {journal} {Appl. Phys.
  Lett.}\ }\textbf {\bibinfo {volume} {97}},\ \bibinfo {pages} {143104}
  (\bibinfo {year} {2010})}\BibitemShut {NoStop}%
\bibitem [{\citenamefont {McGuinness}\ \emph {et~al.}(2011)\citenamefont
  {McGuinness}, \citenamefont {Yan}, \citenamefont {Stacey}, \citenamefont
  {Simpson}, \citenamefont {Hall}, \citenamefont {Maclaurin}, \citenamefont
  {Prawer}, \citenamefont {Mulvaney}, \citenamefont {Wrachtrup}, \citenamefont
  {Caruso}, \citenamefont {Scholten},\ and\ \citenamefont
  {Hollenberg}}]{McguinnessNatNano2011}%
  \BibitemOpen
  \bibfield  {author} {\bibinfo {author} {\bibfnamefont {L.~P.}\ \bibnamefont
  {McGuinness}}, \bibinfo {author} {\bibfnamefont {Y.}~\bibnamefont {Yan}},
  \bibinfo {author} {\bibfnamefont {A.}~\bibnamefont {Stacey}}, \bibinfo
  {author} {\bibfnamefont {D.~A.}\ \bibnamefont {Simpson}}, \bibinfo {author}
  {\bibfnamefont {L.~T.}\ \bibnamefont {Hall}}, \bibinfo {author}
  {\bibfnamefont {D.}~\bibnamefont {Maclaurin}}, \bibinfo {author}
  {\bibfnamefont {S.}~\bibnamefont {Prawer}}, \bibinfo {author} {\bibfnamefont
  {P.}~\bibnamefont {Mulvaney}}, \bibinfo {author} {\bibfnamefont
  {J.}~\bibnamefont {Wrachtrup}}, \bibinfo {author} {\bibfnamefont
  {F.}~\bibnamefont {Caruso}}, \bibinfo {author} {\bibfnamefont {R.~E.}\
  \bibnamefont {Scholten}}, \ and\ \bibinfo {author} {\bibfnamefont {L.~C.~L.}\
  \bibnamefont {Hollenberg}},\ }\href@noop {} {\bibfield  {journal} {\bibinfo
  {journal} {Nat.~Nanotechnol.}\ }\textbf {\bibinfo {volume} {6}},\ \bibinfo
  {pages} {358} (\bibinfo {year} {2011})}\BibitemShut {NoStop}%
\bibitem [{\citenamefont {de~Lange}\ \emph {et~al.}(2011)\citenamefont
  {de~Lange}, \citenamefont {Rist\`e}, \citenamefont {Dobrovitski},\ and\
  \citenamefont {Hanson}}]{deLangePRL2011}%
  \BibitemOpen
  \bibfield  {author} {\bibinfo {author} {\bibfnamefont {G.}~\bibnamefont
  {de~Lange}}, \bibinfo {author} {\bibfnamefont {D.}~\bibnamefont {Rist\`e}},
  \bibinfo {author} {\bibfnamefont {V.~V.}\ \bibnamefont {Dobrovitski}}, \ and\
  \bibinfo {author} {\bibfnamefont {R.}~\bibnamefont {Hanson}},\ }\href
  {\doibase 10.1103/PhysRevLett.106.080802} {\bibfield  {journal} {\bibinfo
  {journal} {Phys. Rev. Lett.}\ }\textbf {\bibinfo {volume} {106}},\ \bibinfo
  {pages} {080802} (\bibinfo {year} {2011})}\BibitemShut {NoStop}%
\bibitem{KagamiPhysica2011} S.~Kagami, Y.~Shikano, and K.~Asahi, Physica E {\textbf{43}}, 761 (2011).
\bibitem [{\citenamefont {Horowitz}\ \emph {et~al.}(2012)\citenamefont
  {Horowitz}, \citenamefont {Alem{\'a}n}, \citenamefont {Christle},
  \citenamefont {Cleland},\ and\ \citenamefont {Awschalom}}]{HorowitzPNAS2012}%
  \BibitemOpen
  \bibfield  {author} {\bibinfo {author} {\bibfnamefont {V.~R.}\ \bibnamefont
  {Horowitz}}, \bibinfo {author} {\bibfnamefont {B.~J.}\ \bibnamefont
  {Alem{\'a}n}}, \bibinfo {author} {\bibfnamefont {D.~J.}\ \bibnamefont
  {Christle}}, \bibinfo {author} {\bibfnamefont {A.~N.}\ \bibnamefont
  {Cleland}}, \ and\ \bibinfo {author} {\bibfnamefont {D.~D.}\ \bibnamefont
  {Awschalom}},\ }\href@noop {} {\bibfield  {journal} {\bibinfo  {journal}
  {Proc. Nat. Acad. Sci.}\ }\textbf {\bibinfo {volume} {109}},\ \bibinfo
  {pages} {13493} (\bibinfo {year} {2012})}\BibitemShut {NoStop}%
\bibitem [{\citenamefont {Hirose}\ \emph {et~al.}(2012)\citenamefont {Hirose},
  \citenamefont {Aiello},\ and\ \citenamefont {Cappellaro}}]{HirosePRA2012}%
  \BibitemOpen
  \bibfield  {author} {\bibinfo {author} {\bibfnamefont {M.}~\bibnamefont
  {Hirose}}, \bibinfo {author} {\bibfnamefont {C.~D.}\ \bibnamefont {Aiello}},
  \ and\ \bibinfo {author} {\bibfnamefont {P.}~\bibnamefont {Cappellaro}},\
  }\href {\doibase 10.1103/PhysRevA.86.062320} {\bibfield  {journal} {\bibinfo
  {journal} {Phys. Rev. A}\ }\textbf {\bibinfo {volume} {86}},\ \bibinfo
  {pages} {062320} (\bibinfo {year} {2012})}\BibitemShut {NoStop}%
\bibitem [{\citenamefont {Pham}\ \emph {et~al.}(2012)\citenamefont {Pham},
  \citenamefont {Bar-Gill}, \citenamefont {Belthangady}, \citenamefont
  {Le~Sage}, \citenamefont {Cappellaro}, \citenamefont {Lukin}, \citenamefont
  {Yacoby},\ and\ \citenamefont {Walsworth}}]{PhamPRB2012}%
  \BibitemOpen
  \bibfield  {author} {\bibinfo {author} {\bibfnamefont {L.~M.}\ \bibnamefont
  {Pham}}, \bibinfo {author} {\bibfnamefont {N.}~\bibnamefont {Bar-Gill}},
  \bibinfo {author} {\bibfnamefont {C.}~\bibnamefont {Belthangady}}, \bibinfo
  {author} {\bibfnamefont {D.}~\bibnamefont {Le~Sage}}, \bibinfo {author}
  {\bibfnamefont {P.}~\bibnamefont {Cappellaro}}, \bibinfo {author}
  {\bibfnamefont {M.~D.}\ \bibnamefont {Lukin}}, \bibinfo {author}
  {\bibfnamefont {A.}~\bibnamefont {Yacoby}}, \ and\ \bibinfo {author}
  {\bibfnamefont {R.~L.}\ \bibnamefont {Walsworth}},\ }\href {\doibase
  10.1103/PhysRevB.86.045214} {\bibfield  {journal} {\bibinfo  {journal} {Phys.
  Rev. B}\ }\textbf {\bibinfo {volume} {86}},\ \bibinfo {pages} {045214}
  (\bibinfo {year} {2012})}\BibitemShut {NoStop}%
\bibitem [{\citenamefont {Nusran}\ and\ \citenamefont
  {Dutt}(2013)}]{NusranPRB2013}%
  \BibitemOpen
  \bibfield  {author} {\bibinfo {author} {\bibfnamefont {N.~M.}\ \bibnamefont
  {Nusran}}\ and\ \bibinfo {author} {\bibfnamefont {M.~V.~G.}\ \bibnamefont
  {Dutt}},\ }\href {\doibase 10.1103/PhysRevB.88.220410} {\bibfield  {journal}
  {\bibinfo  {journal} {Phys. Rev. B}\ }\textbf {\bibinfo {volume} {88}},\
  \bibinfo {pages} {220410} (\bibinfo {year} {2013})}\BibitemShut {NoStop}%
\bibitem [{\citenamefont {Loretz}\ \emph {et~al.}(2013)\citenamefont {Loretz},
  \citenamefont {Rosskopf},\ and\ \citenamefont {Degen}}]{LoretzPRL2013}%
  \BibitemOpen
  \bibfield  {author} {\bibinfo {author} {\bibfnamefont {M.}~\bibnamefont
  {Loretz}}, \bibinfo {author} {\bibfnamefont {T.}~\bibnamefont {Rosskopf}}, \
  and\ \bibinfo {author} {\bibfnamefont {C.~L.}\ \bibnamefont {Degen}},\ }\href
  {\doibase 10.1103/PhysRevLett.110.017602} {\bibfield  {journal} {\bibinfo
  {journal} {Phys. Rev. Lett.}\ }\textbf {\bibinfo {volume} {110}},\ \bibinfo
  {pages} {017602} (\bibinfo {year} {2013})}\BibitemShut {NoStop}%
\bibitem [{\citenamefont {Le~Sage}\ \emph {et~al.}(2013)\citenamefont
  {Le~Sage}, \citenamefont {Arai}, \citenamefont {Glenn}, \citenamefont
  {DeVience}, \citenamefont {Pham}, \citenamefont {Rahn-Lee}, \citenamefont
  {Lukin}, \citenamefont {Yacoby}, \citenamefont {Komeili},\ and\ \citenamefont
  {Walsworth}}]{LeSageNature2013}%
  \BibitemOpen
  \bibfield  {author} {\bibinfo {author} {\bibfnamefont {D.}~\bibnamefont
  {Le~Sage}}, \bibinfo {author} {\bibfnamefont {K.}~\bibnamefont {Arai}},
  \bibinfo {author} {\bibfnamefont {D.~R.}\ \bibnamefont {Glenn}}, \bibinfo
  {author} {\bibfnamefont {S.~J.}\ \bibnamefont {DeVience}}, \bibinfo {author}
  {\bibfnamefont {L.~M.}\ \bibnamefont {Pham}}, \bibinfo {author}
  {\bibfnamefont {L.}~\bibnamefont {Rahn-Lee}}, \bibinfo {author}
  {\bibfnamefont {M.~D.}\ \bibnamefont {Lukin}}, \bibinfo {author}
  {\bibfnamefont {A.}~\bibnamefont {Yacoby}}, \bibinfo {author} {\bibfnamefont
  {A.}~\bibnamefont {Komeili}}, \ and\ \bibinfo {author} {\bibfnamefont
  {R.~L.}\ \bibnamefont {Walsworth}},\ }\href@noop {} {\bibfield  {journal}
  {\bibinfo  {journal} {Nature}\ }\textbf {\bibinfo {volume} {496}},\ \bibinfo
  {pages} {486} (\bibinfo {year} {2013})}\BibitemShut {NoStop}%
\bibitem [{\citenamefont {Geiselmann}\ \emph {et~al.}(2013)\citenamefont
  {Geiselmann}, \citenamefont {Juan}, \citenamefont {Renger}, \citenamefont
  {Say}, \citenamefont {Brown}, \citenamefont {de~Abajo}, \citenamefont
  {Koppens},\ and\ \citenamefont {Quidant}}]{GeiselmannNatureNano2013}%
  \BibitemOpen
  \bibfield  {author} {\bibinfo {author} {\bibfnamefont {M.}~\bibnamefont
  {Geiselmann}}, \bibinfo {author} {\bibfnamefont {M.~L.}\ \bibnamefont
  {Juan}}, \bibinfo {author} {\bibfnamefont {J.}~\bibnamefont {Renger}},
  \bibinfo {author} {\bibfnamefont {J.~M.}\ \bibnamefont {Say}}, \bibinfo
  {author} {\bibfnamefont {L.~J.}\ \bibnamefont {Brown}}, \bibinfo {author}
  {\bibfnamefont {F.~J.~G.}\ \bibnamefont {de~Abajo}}, \bibinfo {author}
  {\bibfnamefont {F.}~\bibnamefont {Koppens}}, \ and\ \bibinfo {author}
  {\bibfnamefont {R.}~\bibnamefont {Quidant}},\ }\href@noop {} {\bibfield
  {journal} {\bibinfo  {journal} {Nat. Nanotechnol.}\ }\textbf {\bibinfo
  {volume} {8}},\ \bibinfo {pages} {175} (\bibinfo {year} {2013})}\BibitemShut
  {NoStop}%
\bibitem [{\citenamefont {Magesan}\ \emph {et~al.}(2013)\citenamefont
  {Magesan}, \citenamefont {Cooper}, \citenamefont {Yum},\ and\ \citenamefont
  {Cappellaro}}]{MagesanPRA2013}%
  \BibitemOpen
  \bibfield  {author} {\bibinfo {author} {\bibfnamefont {E.}~\bibnamefont
  {Magesan}}, \bibinfo {author} {\bibfnamefont {A.}~\bibnamefont {Cooper}},
  \bibinfo {author} {\bibfnamefont {H.}~\bibnamefont {Yum}}, \ and\ \bibinfo
  {author} {\bibfnamefont {P.}~\bibnamefont {Cappellaro}},\ }\href {\doibase
  10.1103/PhysRevA.88.032107} {\bibfield  {journal} {\bibinfo  {journal} {Phys.
  Rev. A}\ }\textbf {\bibinfo {volume} {88}},\ \bibinfo {pages} {032107}
  (\bibinfo {year} {2013})}\BibitemShut {NoStop}%
\bibitem [{\citenamefont {Cooper}\ \emph {et~al.}(2014)\citenamefont {Cooper},
  \citenamefont {Magesan}, \citenamefont {Yum},\ and\ \citenamefont
  {Cappellaro}}]{CooperNatCommun2014}%
  \BibitemOpen
  \bibfield  {author} {\bibinfo {author} {\bibfnamefont {A.}~\bibnamefont
  {Cooper}}, \bibinfo {author} {\bibfnamefont {E.}~\bibnamefont {Magesan}},
  \bibinfo {author} {\bibfnamefont {H.~N.}\ \bibnamefont {Yum}}, \ and\
  \bibinfo {author} {\bibfnamefont {P.}~\bibnamefont {Cappellaro}},\
  }\href@noop {} {\bibfield  {journal} {\bibinfo  {journal} {Nat. Commun.}\
  }\textbf {\bibinfo {volume} {5}},\ \bibinfo {pages} {3141} (\bibinfo {year}
  {2014})}\BibitemShut {NoStop}%
\bibitem [{\citenamefont {Nusran}\ and\ \citenamefont
  {Dutt}(2014)}]{NusranPRB2014}%
  \BibitemOpen
  \bibfield  {author} {\bibinfo {author} {\bibfnamefont {N.~M.}\ \bibnamefont
  {Nusran}}\ and\ \bibinfo {author} {\bibfnamefont {M.~V.~G.}\ \bibnamefont
  {Dutt}},\ }\href {\doibase 10.1103/PhysRevB.90.024422} {\bibfield  {journal}
  {\bibinfo  {journal} {Phys. Rev. B}\ }\textbf {\bibinfo {volume} {90}},\
  \bibinfo {pages} {024422} (\bibinfo {year} {2014})}\BibitemShut {NoStop}%
\bibitem [{\citenamefont {Jensen}\ \emph {et~al.}(2014)\citenamefont {Jensen},
  \citenamefont {Leefer}, \citenamefont {Jarmola}, \citenamefont {Dumeige},
  \citenamefont {Acosta}, \citenamefont {Kehayias}, \citenamefont {Patton},\
  and\ \citenamefont {Budker}}]{JensenPRL2014}%
  \BibitemOpen
  \bibfield  {author} {\bibinfo {author} {\bibfnamefont {K.}~\bibnamefont
  {Jensen}}, \bibinfo {author} {\bibfnamefont {N.}~\bibnamefont {Leefer}},
  \bibinfo {author} {\bibfnamefont {A.}~\bibnamefont {Jarmola}}, \bibinfo
  {author} {\bibfnamefont {Y.}~\bibnamefont {Dumeige}}, \bibinfo {author}
  {\bibfnamefont {V.~M.}\ \bibnamefont {Acosta}}, \bibinfo {author}
  {\bibfnamefont {P.}~\bibnamefont {Kehayias}}, \bibinfo {author}
  {\bibfnamefont {B.}~\bibnamefont {Patton}}, \ and\ \bibinfo {author}
  {\bibfnamefont {D.}~\bibnamefont {Budker}},\ }\href {\doibase
  10.1103/PhysRevLett.112.160802} {\bibfield  {journal} {\bibinfo  {journal}
  {Phys. Rev. Lett.}\ }\textbf {\bibinfo {volume} {112}},\ \bibinfo {pages}
  {160802} (\bibinfo {year} {2014})}\BibitemShut {NoStop}%
\bibitem [{\citenamefont {Chaudhry}(2014)}]{ChaudhryPRA2014NV}%
  \BibitemOpen
  \bibfield  {author} {\bibinfo {author} {\bibfnamefont {A.~Z.}\ \bibnamefont
  {Chaudhry}},\ }\href {\doibase 10.1103/PhysRevA.90.042104} {\bibfield
  {journal} {\bibinfo  {journal} {Phys. Rev. A}\ }\textbf {\bibinfo {volume}
  {90}},\ \bibinfo {pages} {042104} (\bibinfo {year} {2014})}\BibitemShut
  {NoStop}%
\bibitem [{\citenamefont {Hall}\ \emph {et~al.}(2013)\citenamefont {Hall},
  \citenamefont {Simpson},\ and\ \citenamefont {Hollenberg}}]{HallMRS2013}%
  \BibitemOpen
  \bibfield  {author} {\bibinfo {author} {\bibfnamefont {L.}~\bibnamefont
  {Hall}}, \bibinfo {author} {\bibfnamefont {D.}~\bibnamefont {Simpson}}, \
  and\ \bibinfo {author} {\bibfnamefont {L.}~\bibnamefont {Hollenberg}},\
  }\href {\doibase 10.1557/mrs.2013.24} {\bibfield  {journal} {\bibinfo
  {journal} {MRS Bull.}\ }\textbf {\bibinfo {volume} {38}},\ \bibinfo {pages}
  {162} (\bibinfo {year} {2013})}\BibitemShut {NoStop}%
\bibitem [{\citenamefont {Hong}\ \emph {et~al.}(2013)\citenamefont {Hong},
  \citenamefont {Grinolds}, \citenamefont {Pham}, \citenamefont {Le~Sage},
  \citenamefont {Luan}, \citenamefont {Walsworth},\ and\ \citenamefont
  {Yacoby}}]{HongMRS2013}%
  \BibitemOpen
  \bibfield  {author} {\bibinfo {author} {\bibfnamefont {S.}~\bibnamefont
  {Hong}}, \bibinfo {author} {\bibfnamefont {M.~S.}\ \bibnamefont {Grinolds}},
  \bibinfo {author} {\bibfnamefont {L.~M.}\ \bibnamefont {Pham}}, \bibinfo
  {author} {\bibfnamefont {D.}~\bibnamefont {Le~Sage}}, \bibinfo {author}
  {\bibfnamefont {L.}~\bibnamefont {Luan}}, \bibinfo {author} {\bibfnamefont
  {R.~L.}\ \bibnamefont {Walsworth}}, \ and\ \bibinfo {author} {\bibfnamefont
  {A.}~\bibnamefont {Yacoby}},\ }\href {\doibase 10.1557/mrs.2013.23}
  {\bibfield  {journal} {\bibinfo  {journal} {MRS Bull.}\ }\textbf {\bibinfo
  {volume} {38}},\ \bibinfo {pages} {155} (\bibinfo {year} {2013})}\BibitemShut
  {NoStop}%
\bibitem [{\citenamefont {Rondin}\ \emph {et~al.}(2014)\citenamefont {Rondin},
  \citenamefont {Tetienne}, \citenamefont {Hingant}, \citenamefont {Roch},
  \citenamefont {Maletinsky},\ and\ \citenamefont
  {Jacques}}]{RondinRepProgPhys2014}%
  \BibitemOpen
  \bibfield  {author} {\bibinfo {author} {\bibfnamefont {L.}~\bibnamefont
  {Rondin}}, \bibinfo {author} {\bibfnamefont {J.-P.}\ \bibnamefont
  {Tetienne}}, \bibinfo {author} {\bibfnamefont {T.}~\bibnamefont {Hingant}},
  \bibinfo {author} {\bibfnamefont {J.-F.}\ \bibnamefont {Roch}}, \bibinfo
  {author} {\bibfnamefont {P.}~\bibnamefont {Maletinsky}}, \ and\ \bibinfo
  {author} {\bibfnamefont {V.}~\bibnamefont {Jacques}},\ }\href@noop {}
  {\bibfield  {journal} {\bibinfo  {journal} {Rep. Prog. Phys.}\ }\textbf
  {\bibinfo {volume} {77}},\ \bibinfo {pages} {056503} (\bibinfo {year}
  {2014})}\BibitemShut {NoStop}%
\bibitem [{\citenamefont {Helstrom}(1976)}]{Helstrombook}%
  \BibitemOpen
  \bibfield  {author} {\bibinfo {author} {\bibfnamefont {C.~W.}\ \bibnamefont
  {Helstrom}},\ }\href@noop {} {\emph {\bibinfo {title} {Quantum Detection and
  Estimation Theory}}}\ (\bibinfo  {publisher} {Academic},\ \bibinfo {address}
  {New York},\ \bibinfo {year} {1976})\BibitemShut {NoStop}%
\bibitem [{\citenamefont {Holevo}(1982)}]{Holevobook}%
  \BibitemOpen
  \bibfield  {author} {\bibinfo {author} {\bibfnamefont {A.~S.}\ \bibnamefont
  {Holevo}},\ }\href@noop {} {\emph {\bibinfo {title} {Probabilistic and
  Statistical Aspects of Quantum Theory}}}\ (\bibinfo  {publisher}
  {North-Holland},\ \bibinfo {address} {Amsterdam},\ \bibinfo {year}
  {1982})\BibitemShut {NoStop}%
\bibitem [{\citenamefont {Herzog}(2004)}]{HerzogJOpt2004}%
  \BibitemOpen
  \bibfield  {author} {\bibinfo {author} {\bibfnamefont {U.}~\bibnamefont
  {Herzog}},\ }\href@noop {} {\bibfield  {journal} {\bibinfo  {journal} {J.
  Opt. B: Quantum Semiclass. Opt.}\ }\textbf {\bibinfo {volume} {6}},\ \bibinfo
  {pages} {S24} (\bibinfo {year} {2004})}\BibitemShut {NoStop}%
\bibitem [{\citenamefont {Bergou}(2010)}]{BergouOptics2010}%
  \BibitemOpen
  \bibfield  {author} {\bibinfo {author} {\bibfnamefont {J.~A.}\ \bibnamefont
  {Bergou}},\ }\href@noop {} {\bibfield  {journal} {\bibinfo  {journal} {J.
  Mod. Opt.}\ }\textbf {\bibinfo {volume} {57}},\ \bibinfo {pages} {160}
  (\bibinfo {year} {2010})}\BibitemShut {NoStop}%
\bibitem [{\citenamefont {Wang}\ \emph {et~al.}(2012)\citenamefont {Wang},
  \citenamefont {de~Lange}, \citenamefont {Rist\`e}, \citenamefont {Hanson},\
  and\ \citenamefont {Dobrovitski}}]{WanganddeLangePRB2012}%
  \BibitemOpen
  \bibfield  {author} {\bibinfo {author} {\bibfnamefont {Z.-H.}\ \bibnamefont
  {Wang}}, \bibinfo {author} {\bibfnamefont {G.}~\bibnamefont {de~Lange}},
  \bibinfo {author} {\bibfnamefont {D.}~\bibnamefont {Rist\`e}}, \bibinfo
  {author} {\bibfnamefont {R.}~\bibnamefont {Hanson}}, \ and\ \bibinfo {author}
  {\bibfnamefont {V.~V.}\ \bibnamefont {Dobrovitski}},\ }\href {\doibase
  10.1103/PhysRevB.85.155204} {\bibfield  {journal} {\bibinfo  {journal} {Phys.
  Rev. B}\ }\textbf {\bibinfo {volume} {85}},\ \bibinfo {pages} {155204}
  (\bibinfo {year} {2012})}\BibitemShut {NoStop}%
\bibitem [{Note1()}]{Note1}%
  \BibitemOpen
  \bibinfo {note} {Strictly speaking, there are three other cases: one
  eigenvalue positive and the other zero, one eigenvalue negative and the other
  zero, and both eigenvalues zero. It is easy to see that in these cases, we
  can make a decision without performing any measurement.}\BibitemShut {Stop}%
\bibitem [{\citenamefont {Viola}\ and\ \citenamefont
  {Lloyd}(1998)}]{ViolaPRA1998}%
  \BibitemOpen
  \bibfield  {author} {\bibinfo {author} {\bibfnamefont {L.}~\bibnamefont
  {Viola}}\ and\ \bibinfo {author} {\bibfnamefont {S.}~\bibnamefont {Lloyd}},\
  }\href {\doibase 10.1103/PhysRevA.58.2733} {\bibfield  {journal} {\bibinfo
  {journal} {Phys. Rev. A}\ }\textbf {\bibinfo {volume} {58}},\ \bibinfo
  {pages} {2733} (\bibinfo {year} {1998})}\BibitemShut {NoStop}%
\bibitem [{\citenamefont {Viola}\ \emph {et~al.}(1999)\citenamefont {Viola},
  \citenamefont {Knill},\ and\ \citenamefont {Lloyd}}]{LloydPRL1999}%
  \BibitemOpen
  \bibfield  {author} {\bibinfo {author} {\bibfnamefont {L.}~\bibnamefont
  {Viola}}, \bibinfo {author} {\bibfnamefont {E.}~\bibnamefont {Knill}}, \ and\
  \bibinfo {author} {\bibfnamefont {S.}~\bibnamefont {Lloyd}},\ }\href
  {\doibase 10.1103/PhysRevLett.82.2417} {\bibfield  {journal} {\bibinfo
  {journal} {Phys. Rev. Lett.}\ }\textbf {\bibinfo {volume} {82}},\ \bibinfo
  {pages} {2417} (\bibinfo {year} {1999})}\BibitemShut {NoStop}%
\bibitem [{\citenamefont {Biercuk}\ \emph {et~al.}(2009)\citenamefont
  {Biercuk}, \citenamefont {Uys}, \citenamefont {VanDevender}, \citenamefont
  {Shiga}, \citenamefont {Itano},\ and\ \citenamefont
  {Bollinger}}]{BiercukNature2009}%
  \BibitemOpen
  \bibfield  {author} {\bibinfo {author} {\bibfnamefont {M.~J.}\ \bibnamefont
  {Biercuk}}, \bibinfo {author} {\bibfnamefont {H.}~\bibnamefont {Uys}},
  \bibinfo {author} {\bibfnamefont {A.~P.}\ \bibnamefont {VanDevender}},
  \bibinfo {author} {\bibfnamefont {N.}~\bibnamefont {Shiga}}, \bibinfo
  {author} {\bibfnamefont {W.~M.}\ \bibnamefont {Itano}}, \ and\ \bibinfo
  {author} {\bibfnamefont {J.~J.}\ \bibnamefont {Bollinger}},\ }\href {\doibase
  10.1038/nature07951} {\bibfield  {journal} {\bibinfo  {journal} {Nature}\
  }\textbf {\bibinfo {volume} {458}},\ \bibinfo {pages} {996} (\bibinfo {year}
  {2009})}\BibitemShut {NoStop}%
\bibitem [{\citenamefont {Du}\ \emph {et~al.}(2009)\citenamefont {Du},
  \citenamefont {Rong}, \citenamefont {Zhao}, \citenamefont {Wang},
  \citenamefont {Yang},\ and\ \citenamefont {Liu}}]{LiuNature2009}%
  \BibitemOpen
  \bibfield  {author} {\bibinfo {author} {\bibfnamefont {J.}~\bibnamefont
  {Du}}, \bibinfo {author} {\bibfnamefont {X.}~\bibnamefont {Rong}}, \bibinfo
  {author} {\bibfnamefont {N.}~\bibnamefont {Zhao}}, \bibinfo {author}
  {\bibfnamefont {Y.}~\bibnamefont {Wang}}, \bibinfo {author} {\bibfnamefont
  {J.}~\bibnamefont {Yang}}, \ and\ \bibinfo {author} {\bibfnamefont {R.~B.}\
  \bibnamefont {Liu}},\ }\href {\doibase 10.1038/nature08470} {\bibfield
  {journal} {\bibinfo  {journal} {Nature}\ }\textbf {\bibinfo {volume} {461}},\
  \bibinfo {pages} {1265} (\bibinfo {year} {2009})}\BibitemShut {NoStop}%
\bibitem [{\citenamefont {de~Lange}\ \emph {et~al.}(2010)\citenamefont
  {de~Lange}, \citenamefont {Wang}, \citenamefont {Rist\`{e}}, \citenamefont
  {Dobrovitski},\ and\ \citenamefont {Hanson}}]{HansonScience2010}%
  \BibitemOpen
  \bibfield  {author} {\bibinfo {author} {\bibfnamefont {G.}~\bibnamefont
  {de~Lange}}, \bibinfo {author} {\bibfnamefont {Z.~H.}\ \bibnamefont {Wang}},
  \bibinfo {author} {\bibfnamefont {D.}~\bibnamefont {Rist\`{e}}}, \bibinfo
  {author} {\bibfnamefont {V.~V.}\ \bibnamefont {Dobrovitski}}, \ and\ \bibinfo
  {author} {\bibfnamefont {R.}~\bibnamefont {Hanson}},\ }\href {\doibase
  10.1126/science.1192739} {\bibfield  {journal} {\bibinfo  {journal}
  {Science}\ }\textbf {\bibinfo {volume} {330}},\ \bibinfo {pages} {60}
  (\bibinfo {year} {2010})}\BibitemShut {NoStop}%
\bibitem [{\citenamefont {Ryan}\ \emph {et~al.}(2010)\citenamefont {Ryan},
  \citenamefont {Hodges},\ and\ \citenamefont {Cory}}]{RyanPRL2010}%
  \BibitemOpen
  \bibfield  {author} {\bibinfo {author} {\bibfnamefont {C.~A.}\ \bibnamefont
  {Ryan}}, \bibinfo {author} {\bibfnamefont {J.~S.}\ \bibnamefont {Hodges}}, \
  and\ \bibinfo {author} {\bibfnamefont {D.~G.}\ \bibnamefont {Cory}},\ }\href
  {\doibase 10.1103/PhysRevLett.105.200402} {\bibfield  {journal} {\bibinfo
  {journal} {Phys. Rev. Lett.}\ }\textbf {\bibinfo {volume} {105}},\ \bibinfo
  {pages} {200402} (\bibinfo {year} {2010})}\BibitemShut {NoStop}%
\bibitem [{\citenamefont {Naydenov}\ \emph {et~al.}(2011)\citenamefont
  {Naydenov}, \citenamefont {Dolde}, \citenamefont {Hall}, \citenamefont
  {Shin}, \citenamefont {Fedder}, \citenamefont {Hollenberg}, \citenamefont
  {Jelezko},\ and\ \citenamefont {Wrachtrup}}]{NaydenovPRB2011}%
  \BibitemOpen
  \bibfield  {author} {\bibinfo {author} {\bibfnamefont {B.}~\bibnamefont
  {Naydenov}}, \bibinfo {author} {\bibfnamefont {F.}~\bibnamefont {Dolde}},
  \bibinfo {author} {\bibfnamefont {L.~T.}\ \bibnamefont {Hall}}, \bibinfo
  {author} {\bibfnamefont {C.}~\bibnamefont {Shin}}, \bibinfo {author}
  {\bibfnamefont {H.}~\bibnamefont {Fedder}}, \bibinfo {author} {\bibfnamefont
  {L.~C.~L.}\ \bibnamefont {Hollenberg}}, \bibinfo {author} {\bibfnamefont
  {F.}~\bibnamefont {Jelezko}}, \ and\ \bibinfo {author} {\bibfnamefont
  {J.}~\bibnamefont {Wrachtrup}},\ }\href {\doibase 10.1103/PhysRevB.83.081201}
  {\bibfield  {journal} {\bibinfo  {journal} {Phys. Rev. B}\ }\textbf {\bibinfo
  {volume} {83}},\ \bibinfo {pages} {081201} (\bibinfo {year}
  {2011})}\BibitemShut {NoStop}%
\bibitem [{\citenamefont {Bar-Gill}\ \emph {et~al.}(2012)\citenamefont
  {Bar-Gill}, \citenamefont {Pham}, \citenamefont {Belthangady}, \citenamefont
  {Sage}, \citenamefont {Capellaro}, \citenamefont {Maze}, \citenamefont
  {Lukin}, \citenamefont {Yacoby},\ and\ \citenamefont
  {Walsworth}}]{BarGillNatCommun2012}%
  \BibitemOpen
  \bibfield  {author} {\bibinfo {author} {\bibfnamefont {N.}~\bibnamefont
  {Bar-Gill}}, \bibinfo {author} {\bibfnamefont {L.}~\bibnamefont {Pham}},
  \bibinfo {author} {\bibfnamefont {C.}~\bibnamefont {Belthangady}}, \bibinfo
  {author} {\bibfnamefont {D.~L.}\ \bibnamefont {Sage}}, \bibinfo {author}
  {\bibfnamefont {P.}~\bibnamefont {Capellaro}}, \bibinfo {author}
  {\bibfnamefont {J.}~\bibnamefont {Maze}}, \bibinfo {author} {\bibfnamefont
  {M.}~\bibnamefont {Lukin}}, \bibinfo {author} {\bibfnamefont
  {A.}~\bibnamefont {Yacoby}}, \ and\ \bibinfo {author} {\bibfnamefont
  {R.}~\bibnamefont {Walsworth}},\ }\href@noop {} {\bibfield  {journal}
  {\bibinfo  {journal} {Nat. Commun.}\ }\textbf {\bibinfo {volume} {3}},\
  \bibinfo {pages} {858} (\bibinfo {year} {2012})}\BibitemShut {NoStop}%
\bibitem [{\citenamefont {Zhao}\ \emph {et~al.}(2012)\citenamefont {Zhao},
  \citenamefont {Ho},\ and\ \citenamefont {Liu}}]{ZhaoPRB2012}%
  \BibitemOpen
  \bibfield  {author} {\bibinfo {author} {\bibfnamefont {N.}~\bibnamefont
  {Zhao}}, \bibinfo {author} {\bibfnamefont {S.-W.}\ \bibnamefont {Ho}}, \ and\
  \bibinfo {author} {\bibfnamefont {R.-B.}\ \bibnamefont {Liu}},\ }\href
  {\doibase 10.1103/PhysRevB.85.115303} {\bibfield  {journal} {\bibinfo
  {journal} {Phys. Rev. B}\ }\textbf {\bibinfo {volume} {85}},\ \bibinfo
  {pages} {115303} (\bibinfo {year} {2012})}\BibitemShut {NoStop}%
\bibitem [{\citenamefont {Witzel}\ \emph {et~al.}(2012)\citenamefont {Witzel},
  \citenamefont {Carroll}, \citenamefont {Cywi\ifmmode~\acute{n}\else
  \'{n}\fi{}ski},\ and\ \citenamefont {Das~Sarma}}]{WitzelPRB2012}%
  \BibitemOpen
  \bibfield  {author} {\bibinfo {author} {\bibfnamefont {W.~M.}\ \bibnamefont
  {Witzel}}, \bibinfo {author} {\bibfnamefont {M.~S.}\ \bibnamefont {Carroll}},
  \bibinfo {author} {\bibfnamefont {L.}~\bibnamefont
  {Cywi\ifmmode~\acute{n}\else \'{n}\fi{}ski}}, \ and\ \bibinfo {author}
  {\bibfnamefont {S.}~\bibnamefont {Das~Sarma}},\ }\href {\doibase
  10.1103/PhysRevB.86.035452} {\bibfield  {journal} {\bibinfo  {journal} {Phys.
  Rev. B}\ }\textbf {\bibinfo {volume} {86}},\ \bibinfo {pages} {035452}
  (\bibinfo {year} {2012})}\BibitemShut {NoStop}%
\bibitem{supple} For the expressions used to evaluate the effect of decoherence in the case of the CPMG pulse sequence, as well as calculating the effect of non-ideal measurement of the NV center state due to imperfect photon detection, refer to the Supplementary Material.
\bibitem [{\citenamefont {Higgins}\ \emph {et~al.}(2011)\citenamefont
  {Higgins}, \citenamefont {Doherty}, \citenamefont {Bartlett}, \citenamefont
  {Pryde},\ and\ \citenamefont {Wiseman}}]{HigginsPRA2011}%
  \BibitemOpen
  \bibfield  {author} {\bibinfo {author} {\bibfnamefont {B.~L.}\ \bibnamefont
  {Higgins}}, \bibinfo {author} {\bibfnamefont {A.~C.}\ \bibnamefont
  {Doherty}}, \bibinfo {author} {\bibfnamefont {S.~D.}\ \bibnamefont
  {Bartlett}}, \bibinfo {author} {\bibfnamefont {G.~J.}\ \bibnamefont {Pryde}},
  \ and\ \bibinfo {author} {\bibfnamefont {H.~M.}\ \bibnamefont {Wiseman}},\
  }\href {\doibase 10.1103/PhysRevA.83.052314} {\bibfield  {journal} {\bibinfo
  {journal} {Phys. Rev. A}\ }\textbf {\bibinfo {volume} {83}},\ \bibinfo
  {pages} {052314} (\bibinfo {year} {2011})}\BibitemShut {NoStop}%
\end{thebibliography}

\begin{thebibliography}{99}%

\bibitem{suppleWanganddeLangePRB2012} Z.-H.~Wang, G.~de Lange, D.~Rist\`{e}, R.~Hanson, and V.~V.~Dobrovitski, \prb\jn{85}, 155204 (2012).
\bibitem{suppleChaudhryPRA2014NV} A.~Z.~Chaudhry, \pra\jn{90}, 042104 (2014).
\bibitem{suppleRondin2014} L.~Rondin, J.-P.~Tetienne, T.~Hingant, J.-F.~Roch, P.~Maletinsky, and V.~Jacques, Rep.~Prog.~Phys.~\textbf{77}, 056503 (2014).
\bibitem{suppleKennedyNV} See, for instance, T.~A.~Kennedy, F.~T.~Charnock, J.~S.~Colton, J.~E.~Butler, R.~C.~Linares, and P.~J.~Doering, Phys. Status Solidi B \textbf{233}, 416 (2002).


\end{thebibliography}
\end{document}